\documentclass[aps,pra,reprint,superscriptaddress,showpacs]{revtex4-1}
\usepackage[utf8]{inputenc}
\usepackage{graphicx}
\usepackage{multirow}
\usepackage{amsmath,amssymb,amsfonts,amsthm}
\usepackage{xcolor}
\usepackage{braket}
\begin{document}

\title{Quantum Simulation of Spin-Boson Models with Structured Bath}

\author{Ke Sun}
\email{ke.sun@duke.edu}
\affiliation{Duke Quantum Center, Duke University, Durham, NC 27701, USA}
\affiliation{Department of Physics, Duke University, Durham, NC 27708, USA}

\author{Mingyu Kang}
\email{mingyu.kang@duke.edu}
\affiliation{Duke Quantum Center, Duke University, Durham, NC 27701, USA}
\affiliation{Department of Physics, Duke University, Durham, NC 27708, USA}

\author{Hanggai Nuomin}
\email{hanggai.nuomin@duke.edu}
\affiliation{Department of Chemistry, Duke University, Durham, NC 27708, USA}

\author{George Schwartz}
\affiliation{Duke Quantum Center, Duke University, Durham, NC 27701, USA}
\affiliation{Department of Physics, Duke University, Durham, NC 27708, USA}

\author{David N. Beratan}
\affiliation{Duke Quantum Center, Duke University, Durham, NC 27701, USA}
\affiliation{Department of Physics, Duke University, Durham, NC 27708, USA} 
\affiliation{Department of Chemistry, Duke University, Durham, NC 27708, USA}
\affiliation{Department of Biochemistry, Duke University, Durham, NC 27710, USA}

\author{Kenneth R. Brown}
%\email{kenneth.r.brown@duke.edu}
\affiliation{Duke Quantum Center, Duke University, Durham, NC 27701, USA}
\affiliation{Department of Physics, Duke University, Durham, NC 27708, USA}
\affiliation{Department of Chemistry, Duke University, Durham, NC 27708, USA}
\affiliation{Department of Electrical and Computer Engineering, Duke University, Durham, NC 27708, USA}

\author{Jungsang Kim}
%\email{jungsang@duke.edu}
\affiliation{Duke Quantum Center, Duke University, Durham, NC 27701, USA}
\affiliation{Department of Physics, Duke University, Durham, NC 27708, USA}
\affiliation{Department of Electrical and Computer Engineering, Duke University, Durham, NC 27708, USA}

\date{\today}

\begin{abstract}
The spin-boson model, involving spins interacting with a bath of quantum harmonic oscillators, is a widely used representation of open quantum systems that describe many dissipative processes in physical, chemical and biological systems. Trapped ions present an ideal platform for simulating the quantum dynamics of such models, by accessing both the high-quality internal qubit states and the motional modes of the ions for spins and bosons, respectively. We demonstrate a fully programmable method to simulate dissipative dynamics of spin-boson models using a chain of trapped ions, where the initial temperature and the spectral densities of the boson bath are engineered by controlling the state of the motional modes and their coupling with qubit states. Our method provides a versatile and precise experimental tool for studying open quantum systems. \\

A new method to simulate open quantum systems made of spins and harmonic oscillators where the bath properties are engineered.
\end{abstract}

\maketitle

Interaction of a quantum system with its environment in an open dissipative setting can significantly influence its dynamics. A paradigmatic model of non-Markovian open quantum systems that captures the quantum nature of both the system and the environment is the spin-boson model~\cite{Leggett87}, which is ubiquitous in the study of condensed-matter physics~\cite{Caldeira83, Weiss87}, chemical reactions~\cite{Garg85, Gilmore05}, collective modes in atomic-photonic systems~\cite{StrackPRL2011,BhaseenPRA2012}, and biological light-harvesting complexes~\cite{Huelga13, Jang18}. Experimentally realizing the dynamics of spin-boson models in a programmable fashion is an exciting challenge~\cite{Mostame12, Leppakangas18, Magazzu18, Kim22}, which can lead to new quantum simulation approaches that reach beyond classically tractable limit. 

Trapped ions provide a highly controllable system of spins and bosonic motional modes, and thus are a natural platform for simulating spin-boson models. The bosonic bath can be simulated using either a large number ($\gtrsim 50$) of motional modes~\cite{Porras08} or a handful of dissipative motional modes~\cite{Lemmer18, Schlawin21, MacDonell21, Kang24}. While significant experimental progress has been made in controlling the qubits and motional modes together~\cite{Chen23,Fluehmann2020motion,Jia22}, past works in this direction are limited to simulating a spin coherently coupled to up to two bosonic modes~\cite{Whitlow23, Valahu23, Gorman18, Macdonell23}. These studies could not consider the effects of dissipation often described by coupling of the system to a \textit{continuum} of bath modes, which determines the system's long-term behavior such as reaching a steady state. 

Dissipative couplings come in two flavors: {\em damping} is when the quantum system exchanges energy with the bath (inelastic coupling) and eventually reaches thermal equilibrium with the bath, while {\em dephasing} is when the phase of the quantum system is randomized by interaction with the bath without energy exchange (elastic coupling). In this work, we perform quantum simulations of spin-boson models with programmable spectral densities using a 7-ion chain and up to 3 of its motional modes. The target spectral densities of the bath are decomposed into several Lorentzian peaks, which is characteristic of baths in real-world systems such as light-harvesting protein complexes~\cite{Lee16, Kim18}. Our approach can readily simulate the dephasing model of dissipation, by adding randomness to the control parameters of the laser that drives the spin-boson interaction, such as frequency and phase. Using this technique, we study the dynamics of classic spin-boson models~\cite{Leggett87} and the vibration-assisted energy transfer (VAET) process~\cite{Gorman18}.
Our demonstration to engineer spectral density of the bath in analog quantum simulations is an important ingredient for achieving quantum advantage in simulating open quantum system models that are intractable using classical methods~\cite{MacDonell21, Kang24}. 

%Furthermore, we simulate the sub-Ohmic, Ohmic, and super-Ohmic spin-boson models~\cite{Leggett87}, which are ubiquitous in various open quantum systems ranging from nanomechanical devices~\cite{Galland08} and superconducting circuits~\cite{Magazzu18, Kaur21} to molecules in which proton-transfer reactions occur~\cite{Ohta06, Craig07}. Finally, we simulate the vibration-assisted energy transfer (VAET)~\cite{Gorman18} in a simple molecular model where dephasing is added to the vibrational mode. The various spin-state population dynamics, which display behaviors such as damped oscillations, coherent revivals, and relaxation to a steady state, are accurately captured by our quantum-simulation results.

We note that a recent study~\cite{Wang24} simulated a spin dissipatively coupled to many modes, albeit with limited tunability. 

%Capturing the dissipative behavior of the spin-boson model is enabled by adding randomness to the control parameters, such as the laser frequency and phase. Remarkably, the impact of these control signals is directly analogous to motional dephasing, the leading source of noise in our experimental setup~\cite{Kang23}. This serves as a lower limit for the level of noise that can be introduced in our quantum simulation process. 

%\section*{Results}
\section*{Dephased spin-oscillator model}

The spin-boson model describes a spin coupled to a continuous bath of quantum harmonic oscillators. The Hamiltonian is given by 
\begin{equation} \label{eq:spinbosonmodel}
    \hat{H} = \hat{H}_S + \hat{H}_B + \frac{\hat{\sigma}_Z}{2} \otimes \int_0^\infty d\omega \sqrt{\frac{J(\omega)}{\pi}} \left( \hat{a}(\omega) + \hat{a}^\dagger(\omega) \right), 
\end{equation}
where $J(\omega)$ is the spectral density of the bath, $\hat{a}(\omega)$ is the annihilation operator of the bath oscillator of frequency $\omega$, and $\hat{H}_S = \frac{\epsilon}{2} \hat{\sigma}_Z + \frac{\Delta}{2} \hat{\sigma}_X$ and $\hat{H}_B = \int_0^\infty d\omega \: \omega \hat{a}^\dagger(\omega) \hat{a}(\omega)$ are the Hamiltonian of the spin and the bath, respectively. Here, $\epsilon$ and $\Delta$ are the detuning and coupling strength between the spin states, respectively, and $\hbar$ is set to $1$ for simplicity. We assume that initially, the spin is in the $\ket{0}$ state ({\em ``donor''} state), and each bath oscillator of frequency $\omega$ is in the thermal state of average phonon number $\bar{n}(\omega) = 1/(e^{\beta \omega} - 1)$, where $\beta$ is the inverse of the temperature ($k_B = 1$ for simplicity). 

We decompose the spectral density $J(\omega)$ of a structured bath into a sum of multiple Lorentzian peaks, each of which can be represented by a dissipative harmonic oscillator~\cite{Lemmer18}. In this work, we consider a spin coupled to several oscillators subject to constant dephasing ({\em ``dephased''} spin-oscillator model). This ``discrete'' oscillator model is described by the Hamiltonian
\begin{equation} \label{eq:HSO}
    \hat{H}_{D} = \hat{H}_S + \sum_l \frac{\kappa_l}{2} \hat{\sigma}_Z \otimes (\hat{b}_l + \hat{b}_l^\dagger) + \sum_l \nu_l \hat{b}_l^\dagger \hat{b}_l
\end{equation}
and corresponding Lindblad operators $\hat{L}_l = \sqrt{\Gamma}_l \hat{b}_l^\dagger \hat{b}_l$, where $\kappa_l$ represents the coupling strength between the spin and $l$-th oscillator, $\nu_l$ and $\Gamma_l$ denote the frequency and the dephasing rate of the $l$-th oscillator, respectively, and $\hat{b}_l$ is the annihilation operator of the $l$-th oscillator. The composite state $\rho$ of spin and oscillators follows the Lindblad master equation $\dot{\rho} = -i [\hat{H}_{D}, \rho] + \sum_l (\hat{L}_l \rho \hat{L}_l^\dagger - \frac{1}{2} \{ \hat{L}_l^\dagger \hat{L}_l, \rho \})$. In a reasonable regime of parameters ($\Gamma_l < \nu_l/2$, $\beta \ll 2\pi (\Gamma_l/2)^{-1}$) \cite{Lemmer18}, this bath of dephased oscillators is assigned a spectral density composed of Lorentzian peaks~\cite{supp}
\begin{equation}
    J_{\rm Lo}(\omega) = \sum_l \kappa_l^2 \Big( \frac{\Gamma_l/2}{(\Gamma_l/2)^2 + (\omega - \nu_l)^2} - \frac{\Gamma_l/2}{(\Gamma_l/2)^2 + (\omega + \nu_l)^2} \Big), \label{eq:Lorentzian}
\end{equation}
where we assume that  each oscillator is initially in the thermal state with average phonon number $\bar{n}(\nu_l)$.

The spectral density of the bath determines the dynamics of the spin to leading order in $\bar{\kappa} T$, where $\bar{\kappa}$ is the average coupling strength and $T$ is the evolution time~\cite{supp}. Therefore, the spin dynamics of the dephased spin-oscillator model approximately match those of the spin-boson model with $J(\omega) = J_{\rm Lo}(\omega)$ in the weak-$\bar{\kappa}$ regime. In the unbiased ($\epsilon=0$) spin-boson model~\cite{Leggett87, Wang08, Duan17, Strathearn18}, the equilibrium donor-state population is $1/2$ for both models, and we expect the population dynamics of the two models show an excellent match at all times for weak spin-bath coupling. For the case of biased spin-boson models in the stronger coupling regime ($\epsilon \neq 0$, such as discussed in the VAET case), the dynamics of the dephased spin-oscillator model we simulate can deviate significantly from the traditional spin-boson model. 

\begin{figure}[ht!]
\includegraphics[width=0.48\textwidth]{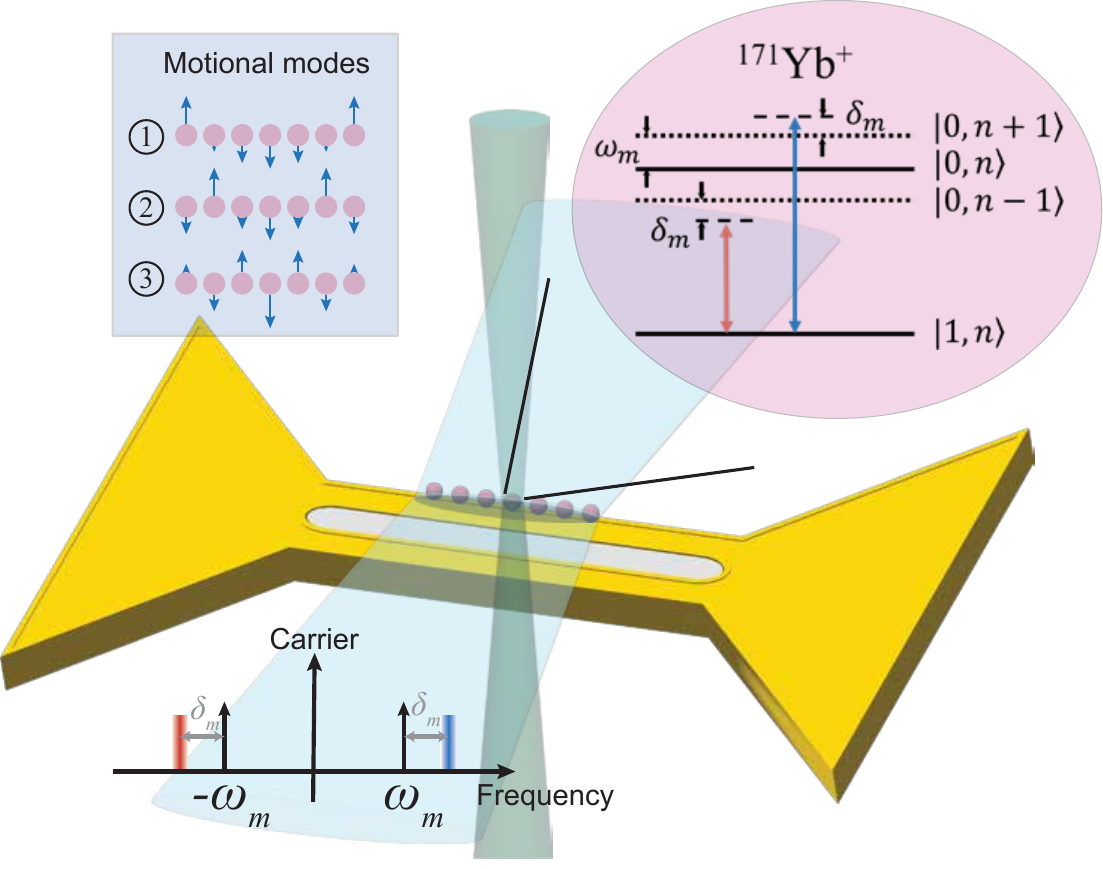}
\caption{\textbf{Schematic of the experimental setup.}  The ions are confined in a micro-fabricated surface trap. The spin states are encoded as the hyperfine clock states (qubit states) and the bosonic bath modes are encoded as the collective radial motional modes of the ion chain. We strategically use the center ion and the three symmetric modes (except for the center-of-mass mode, as shown in the upper left corner) in the experiment to minimize cross-mode couplings~\cite{supp}. The laser frequencies for the SDK operations are determined by the frequency difference between the two Raman beams, and are depicted by the red and blue lines on the frequency axis detuned by $\delta_m$ from the mode frequency $\omega_m$, with the color gradients indicating the random frequencies used to implement the decoherence simulation.} \label{fig_experiment}
\end{figure}

A more common model for dissipation is a spin coupled to oscillators subject to constant damping ({\em ``damped''} spin-oscillator model) described by a pair of Lindblad operators $\hat{L}_{l1} = \sqrt{\Gamma_l (\bar{n}(\nu_l)+1)} \hat{b}_l$ and $\hat{L}_{l2} = \sqrt{\Gamma_l \bar{n}(\nu_l)} \hat{b}_l^\dagger$~\cite{Lemmer18, Schlawin21}, where $\Gamma_l$ here is the damping rate of the $l$-th oscillator. A stronger result holds for this model: the spin dynamics are non-perturbatively equivalent to those of the spin-boson model with the same spectral density, up to all orders in $\bar{\kappa} T$~\cite{Tamascelli18, supp}. Damped oscillators can be realized in trapped ions using sympathetic cooling on a chain of ions with multiple atomic species or isotopes~\cite{Lemmer18, Schlawin21}. The dephased oscillator model developed in this work expands the ability to simulate a broader set of bath models not experimentally accessible before.

\section*{Experimental Implementation}

\begin{figure*}[ht!]
\includegraphics[width=1.0\textwidth]{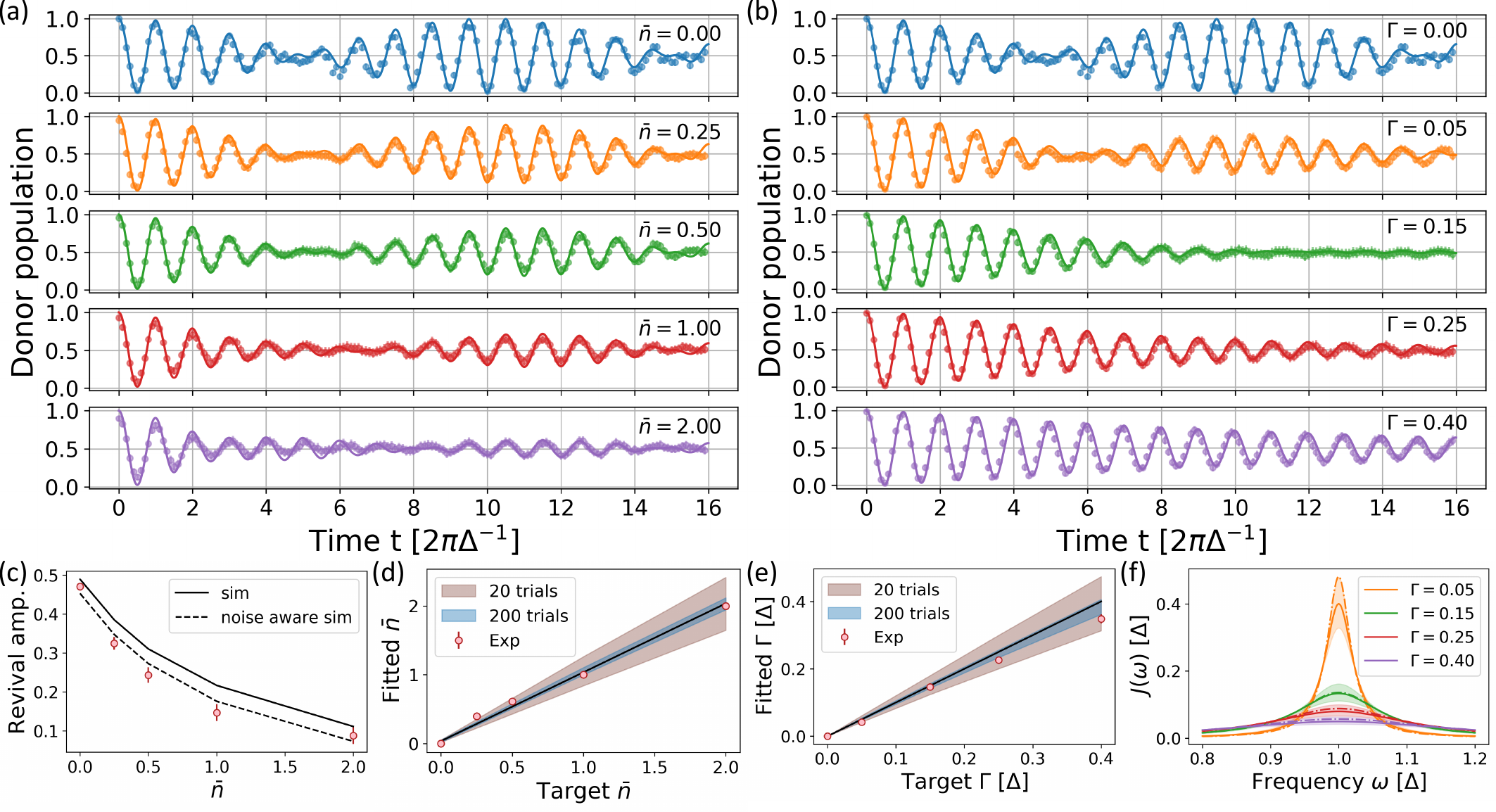}
\caption{\textbf{Demonstration of the tunability of the bath's initial temperature and spectral linewidth.} \textbf{(a)} Dynamics of the coherent spin-single oscillator model $\hat{H}_{D}$ with various values of average phonon number $\bar{n}$ for the oscillator's initial state. \textbf{(b)} Dynamics of the model with $\bar{n}=0$, $J(\omega) = J_{\rm Lo}(\omega)$ (single peak) and various values of $\Gamma_1 \equiv \Gamma$. For both (a) and (b), $\epsilon=0$, $\nu_1=\Delta$, and $\kappa_1=0.1\Delta$. Lines and dots represent theoretical predictions and experimental data, respectively. The error bars (size comparable to the symbols in these two plots) denote measured standard deviation over 20 trials, each trial performed with randomly drawn set of parameters and repeated 100 times. \textbf{(c)} Revival amplitude versus $\bar{n}$. Predictions of the original (solid line) model and the noise-aware (dashed line) model are compared with the experimental results (circles). \textbf{(d)} Measured values of $\bar{n}$ fitted to the predictions of the noise-aware model. Solid line depicts $y = x+\bar{n}_0$, where $\bar{n}_0 = 0.036$ is the expected offset in the phonon number due to imperfect cooling. \textbf{(e)} Measured values of $\Gamma$ fitted to the theoretical predictions with $\bar{n} = \bar{n}_0$. Solid line depicts $y=x$. In (d) and (e), the blue and brown shaded regions represent the expected uncertainty of the value when the populations are averaged over 20 and 200 random trials, respectively. \textbf{(f)} Lorentzian spectral densities for various target (solid) and fitted (dashed) values of $\Gamma$. The shaded region represents the expected uncertainty of $\Gamma$ when averaged over 20 random trials.}\label{fig_tunability}
\end{figure*}

The simulations of spin-boson models are performed using a linear chain of seven $^{171}$Yb$^+$ ions (Figure \ref{fig_experiment}). Two hyperfine internal states ($\ket{0}$ and $\ket{1}$) are used to represent the spin (or qubit). The ions sit 68 $\mu$m above the surface of the trap~\cite{Revelle20}, and the heating rate and decoherence time of the zig-zag motional mode are $3.6(3)$ quanta/s and $5.2(7)$ ms, respectively. The other motional modes used have similar magnitudes. Details of our experimental setup can be found in Ref.~\cite{Wang20}. Using Doppler, electromagnetically-induced transparency (EIT), and sideband cooling techniques, the zig-zag motional mode can be cooled to near ground state with $\bar{n} = 0.036(16)$. Standard qubit manipulation techniques can be used to initialize and measure the qubits, and single-qubit operations are driven by stimulated Raman transitions. The spin-oscillator coupling is simulated using the spin-dependent kick (SDK) operation induced by simultaneous application of blue- and red-sideband transitions~\cite{supp}. 

The evolution of $\hat{H}_{D}$ in Eq.~\eqref{eq:HSO} is mapped to a sequence of single-qubit rotations and SDK operations via Trotterization. We apply time evolution operator over a short time interval corresponding to each term in the Hamiltonian in interaction picture, and repeat them over many time intervals to simulate the evolution dynamics~\cite{supp}. This approach can be readily extended to simulating more complicated molecular models consisting of many electronic states and bath modes~\cite{Sun23,Kang24}.

The dephased spin-oscillator model can be simulated using trapped ions by applying the SDK operations, which induce the coupling between the qubit and the motional mode. By adding randomness to the control parameters of the SDK operations and averaging the results over many random trials, we can implement dissipative processes such as preparing the thermal state and simulating the dephasing of the motional mode. The key idea is that certain dissipative evolutions described by the Lindblad master equation are equivalent to an average of many coherent evolutions, each subject to a stochastic Hamiltonian~\cite{Chenu17}. Details of the procedure is described in Supplementary Material~\cite{supp}.

\section*{Single-oscillator case}

We first consider a single-oscillator case ($l=1$), and demonstrate the programmability of the bath's initial temperature quantified by the oscillator's average phonon number $\bar{n}$. Controlled amounts of resonant SDK operations with stochastically varying phase are applied to the ion prepared near the motional ground state, which prepares the thermal state equivalent to an ensemble of randomly displaced coherent states~\cite{supp}. Figure \ref{fig_tunability}a illustrates the simulated time evolution of the donor state population when it is coupled to such initial bath state, for various values of $\bar{n}$. The dynamics exhibit coherent oscillations with an envelope of slow collapse and revival (``beat-note'') that signifies resonant ($\nu_1=\Delta$) energy exchange between the system and the bath oscillator~\cite{Eberly80}. A thermal state of larger $\bar{n}$ is a mixture of more ``scattered'' coherent states, leading to reduced beat-note amplitude at \mbox{$t \approx 2\pi/\Delta \times 10.5$}. %We see a notable agreement between theoretical predictions (solid lines) and experimental results (dots). 

We compare our experimental results with the predictions of \textit{noise-aware} models, reflecting the noise processes in the experimental setup. We fit the experimental data using an oscillator model with initial phonon number $\bar{n} + \bar{n}_0$ and Lorentzian spectral density with full width at half maximum (FWHM) $\Gamma_1 \equiv \Gamma$ [Eq.~\eqref{eq:Lorentzian}], where the values $\bar{n}_0 = 0.036$ and \mbox{$\Gamma=0.0022\Delta$} are extracted from independent measurements characterizing imperfect cooling and finite motional coherence time, respectively. The experimentally measured revival amplitude (Fig.~\ref{fig_tunability}c) and fitted $\bar{n}$ (Fig.~\ref{fig_tunability}d) match well with the predictions of noise-aware models. This shows that well-characterized experimental noise can serve as the baseline values for the thermal excitation and dephasing rate in the open quantum system model, and thus is not always a bottleneck for the accuracy of our simulations. 

Next, we show the tunability of the dephasing rate $\Gamma$, which corresponds to the linewidth of the spin-boson model's spectral density. Dephasing is implemented by adding random frequency offsets to the SDK operations that simulate the time evolution~\cite{supp}. This mimics the random fluctuations of the motional-mode frequency, following the stochastic model of motional dephasing noise in trapped-ion systems~\cite{Wang20}. 
Figure~\ref{fig_tunability}b illustrates the dynamics for various values of $\Gamma$ between 0 and $0.4\Delta$, which exhibit a crossover from underdamped to overdamped dynamics as $\Gamma$ is increased (critical damping condition at $\Gamma \approx 0.18\Delta$). As the damping is increased beyond the critical point, the beat-note that signifies the excitation transfer from the spin to boson back to spin is suppressed, and there is a monotonic decrease in the oscillation of the donor population in the overdamped regime. We notice a $\pi$ phase flip in the oscillation of the donor population in the beat-note in the underdamped case, signified by the phase difference between the underdamped and overdamped cases at  $t \approx 2\pi/\Delta \times 10$. % These features match very well with the theoretical predictions of spin-boson models. 

\begin{figure}[ht!]
\includegraphics[width=0.48\textwidth]{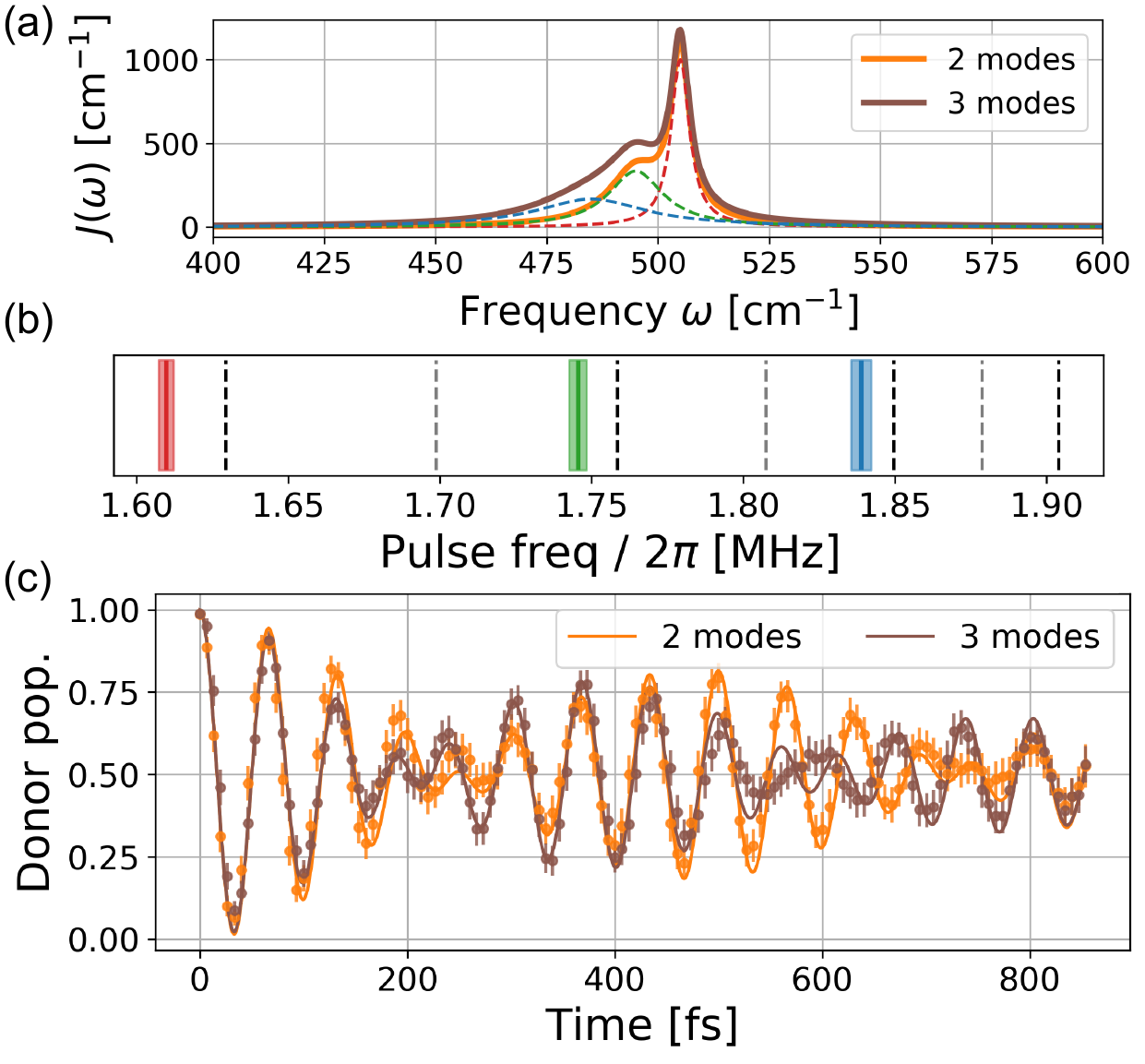}
\caption{\textbf{Simulations of molecular energy transfer in model structured baths.} \textbf{(a)} Simulated spectral density of the spin-boson model's bath, where the electronic coupling strength \mbox{$\Delta=500\:\text{cm}^{-1}$}. The orange and brown solid curves illustrate the spectral densities summed over two and three peaks, respectively. \textbf{(b)} Dashed lines represent the radial motional modes that are coupled by the Raman beam. %The relative ion-mode coupling strengths can be found in Methods.
Black or grey dashed lines represent modes with non-zero or zero coupling strength to the target ion, respectively. The three solid lines and the shaded regions depict the mean value and the standard deviation of the randomized pulse frequencies, and the color corresponds to the Lorentzian peak generating the final spectral density in (a). \textbf{(c)} Time evolution of the spin-boson models with $J(\omega)$ in (a). The bath temperature is \mbox{77 K} such that $\bar{n} \approx 0.1$, which is set for trapped ions' motional modes using imperfect sideband cooling. Curves and dots represent theoretical predictions and experimental data, respectively. Error bars are derived as in Fig.~\ref{fig_tunability}.}\label{fig_3mode}
\end{figure}

The uncertainty of $\bar{n}$ and $\Gamma$ realized in the experiment can be reduced by performing a larger number of random trials, each trial using a different set of random parameters. The shaded areas in Fig.~\ref{fig_tunability}d and e show the standard deviation of fitted values for $\bar{n}$ and $\Gamma$, obtained by numerical simulation of results averaged over 20 (pink) or 200 (blue) random trials. Our experiments are limited to 20 random trials due to the compiling delay required for each set of control parameters, and each trial is repeated 100 times to obtain the expectation value for the donor  population. The experimental data is consistent with the standard deviation derived from 20 numerical trials. We expect to readily increase the number of random trials with fewer repetitions each in future experiments with faster compilation tools~\cite{Dalvi23,supp}. 

\section*{Engineering bath spectral densities}

\begin{figure*}[ht!]
\includegraphics[width=0.95\textwidth]{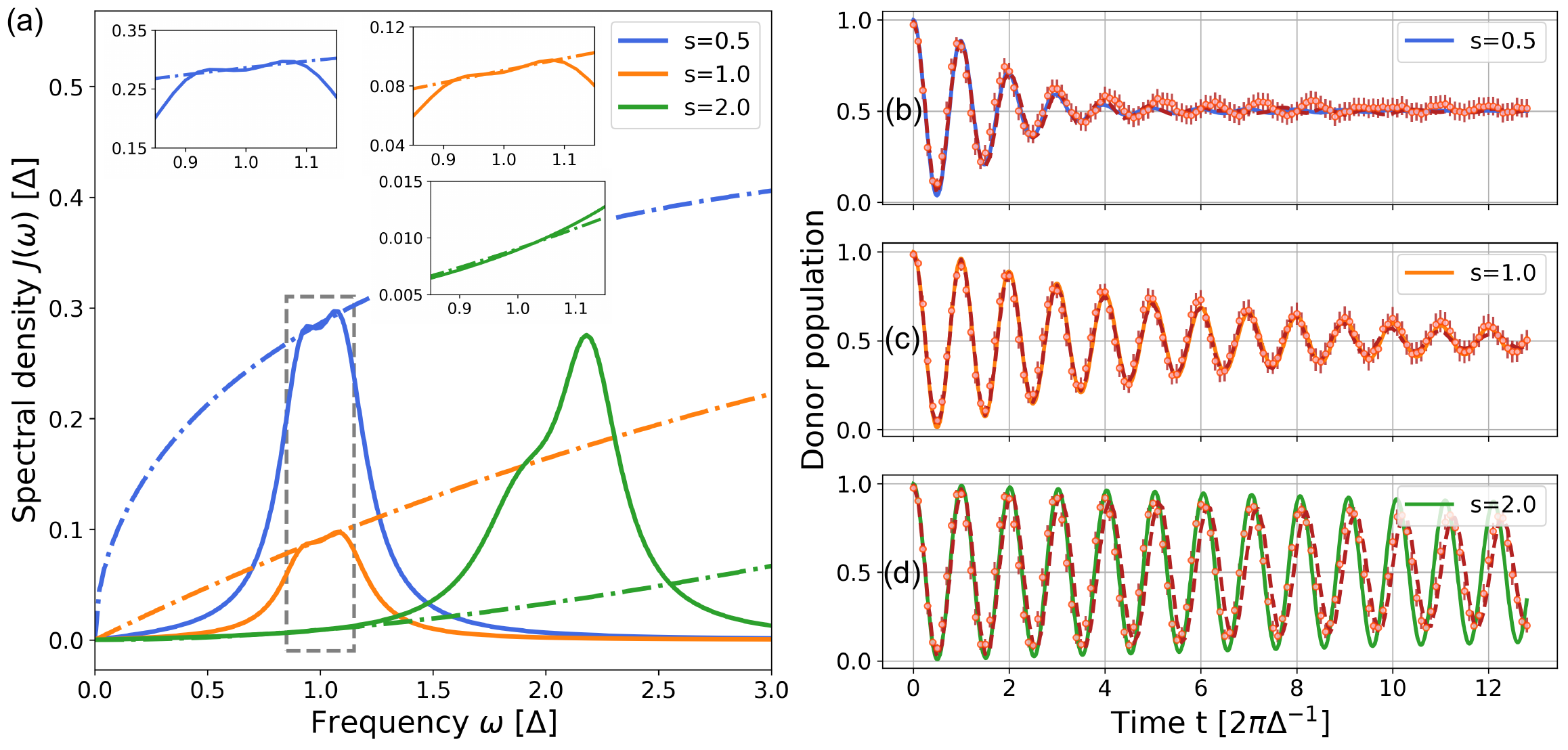}
\caption{\textbf{Simulations based on Leggett's spin-boson model.} \textbf{(a)} Spectral densities of the Leggett's model. Dot-dashed curves are the sub-Ohmic ($s=0.5$), Ohmic ($s=1.0$), and super-Ohmic ($s = 2.0$) spectral densities described by Eq.~\eqref{eq:leggett} ($A=0.1$, $\omega_c = 10\Delta$). Solid curves are the spectral densities simulated in our experiments using two motional modes. These spectral densities are designed to match the dot-dashed curves near $\omega = \Delta$. Insets compare the solid and dot-dashed curves within the interval $\omega \in [0.85\Delta, 1.15\Delta]$. \textbf{(b)} \textbf{(c)} \textbf{(d)} Dynamics of the spin-boson model for $s = 0.5$, $1.0$, and $2.0$, respectively. The solid curves represent the theoretical predictions of the spin-boson models with spectral densities given by the solid curves in (a). The red dashed curves are expected results, derived from numerical simulations of the density matrix of the qubit and motional modes subject to control operations and hardware noise. Circles denote experimental data and error bars indicate the standard deviation over 60, 20, and 20 trials for (b), (c), and (d), respectively. 
}\label{fig_Leggett}
\end{figure*}

We can use multiple oscillator modes to engineer a target spectral density structure for the oscillator bath. We simulate a bath of spectral density composed of up to 3 Lorentzian peaks, each represented by a motional mode of the ion chain. The bath spectral density $J(\omega)$ in Fig.~\ref{fig_3mode}a is simulated using the SDK operations over two or three motional modes. Figure~\ref{fig_3mode}b shows the frequencies of the laser pulses used to drive the SDK, defined as the detuning of laser beams from the carrier transition frequency. The 1st, 3rd, and 5th motional modes are used, and the detuning of the laser frequency from each motional mode determines the frequency at which the bath spectral density is considered in  Fig.~\ref{fig_3mode}a. We set the electronic coupling strength \mbox{$\Delta = 500$ cm$^{-1}$}~\cite{Chenu15}, such that at temperature \mbox{77 K}, $\bar{n} \approx 0.1$ for the near-resonance vibrational modes.

Figure~\ref{fig_3mode}c shows the results for simulating the dynamics of spin-boson models with spectral densities composed of 2 and 3 Lorentzian peaks, respectively. The theoretical predictions are obtained using the time-dependent density-matrix renormalization group algorithm in the interaction picture~\cite{Nuomin22}. The donor population features underdamped coherent oscillations, with contributions from all three modes. The measured population closely follows the theoretical predictions.

The coherence time of the motional modes in our setup of $\sim 5.2(7)$ ms is the leading source of noise~\cite{Wang20,Kang23}, and is comparable to the experimental time for simulating the time evolution under spectral densities generated from 2 mode (1.63 ms) and 3-mode (2.56 ms) cases. For these simulations, the contribution of motional decoherence in the experiment is identical to the random kicks that we apply in the spin-oscillator model to simulate the bath, and can be incorporated as the source of dephasing in our simulation with proper calibration~\cite{Kang23}.

\section*{Leggett spin-boson models}

We further simulate the spin-boson models inspired by Leggett \textit{et al.}~\cite{Leggett87}. The spectral density of the bath is given by
\begin{equation} 
    J_{\rm Legg}(\omega) = A \omega^s \omega_c^{1-s} e^{-\omega / \omega_c}, \label{eq:leggett}
\end{equation}
where $\omega_c$ is the cutoff frequency, and $s<1$, $s=1$, and $s>1$ represent the sub-Ohmic, Ohmic and super-Ohmic baths, respectively. This model is widely used to characterize the noise in nanomechanical devices~\cite{Galland08}, superconducting circuits~\cite{Magazzu18, Kaur21}, and proton-transfer reactions~\cite{Ohta06, Craig07}. The spin dynamics are expected to exhibit a phase transition from coherent oscillations to incoherent decay or localization as $s$ is decreased or the amplitude $A$ is increased, which has been a subject of longstanding research using analytical tools~\cite{Leggett87, Bulla03} and classical simulations~\cite{Wang08, Duan17, Strathearn18, Nalbach13}. 

Recognizing that the spin exchanges energy with the bath only near its resonance, we use a sum of Lorentzian lines from motional modes to approximate the spectral density $J_{\rm Legg}(\omega)$ near the resonance frequency $\Delta$. We utilize well-established global optimization algorithms to find the optimal set of parameters ($\kappa_l$, $\Gamma_l$, and $\nu_l$ in Eq.~\eqref{eq:Lorentzian}) that best represent $J_{\rm Legg}(\omega)$~\cite{supp}. Using this approach, we consider weak spin-bath coupling ($A=0.1$) and match the spectral density within the target bandwidth $\omega \in [0.9\Delta, 1.1\Delta]$ with 2 Lorentzian peaks. Figure~\ref{fig_Leggett}a shows the obtained spectra from the modes (solid lines) that approximate sub-Ohmic \mbox{($s=0.5$)}, Ohmic \mbox{($s=1.0$)}, and super-Ohmic \mbox{($s=2.0$)} spectral densities (dot-dashed lines), respectively. The results for the donor population dynamics interacting with a bath featuring these approximated spectral densities are shown in Fig.~\ref{fig_Leggett}b-d. We observe a crossover from near-coherent ($s=2$) to fully damped ($s=0.5$) oscillations in the simulated time scale, owing to the different  levels of spectral density $J(\omega=\Delta)$ near resonance. The observed behavior qualitatively agrees with the predictions of the Leggett model: for super-Ohmic case, coherence is maintained due to low spectral density near resonance, while for sub-Ohmic case, large spectral density induces strong damping to the point where the system is localized. We note that the deviations in the $s=2$ case in Fig.~\ref{fig_Leggett}d result from insufficient number of Trotterization steps~\cite{supp}.

%Figure~\ref{fig_Leggett}e compares the predicted population dynamics of the spin-boson model with ideal spectral density  $J_{\rm Legg}(\omega)$ and the corresponding approximated spectral density. The dynamics match very well for the Ohmic and super-Ohmic cases; however, the sub-Ohmic case shows a non-negligible deviation due to the limited bandwidth. While we focus on accurately capturing a narrow bandwidth in line with the physical assumptions in the Leggett model, our method can be extended to matching a wider bandwidth in a relatively strong spin-bath coupling regime, which can be useful in studying other spin-boson models~\cite{supp}. 
\section*{Vibration-assisted energy transfer}

Finally, we simulate the VAET model~\cite{Gorman18} with nonzero energy detuning $\epsilon$ between the spin states. Here we set the spin parameters as \mbox{$\Delta = 30\:\text{cm}^{-1}$} and \mbox{$\epsilon = 100\:\text{cm}^{-1}$}. In the absence of coupling to a vibrational mode ($\kappa=0 \:\text{cm}^{-1}$), energy transfer between the detuned spin states is suppressed; however, for nonzero $\kappa$, the energy transfer can be activated if the mode frequency $\nu$ meets the resonance condition (single mode's index $l=1$ omitted).

\begin{figure}[ht!]
\includegraphics[width=0.48\textwidth]{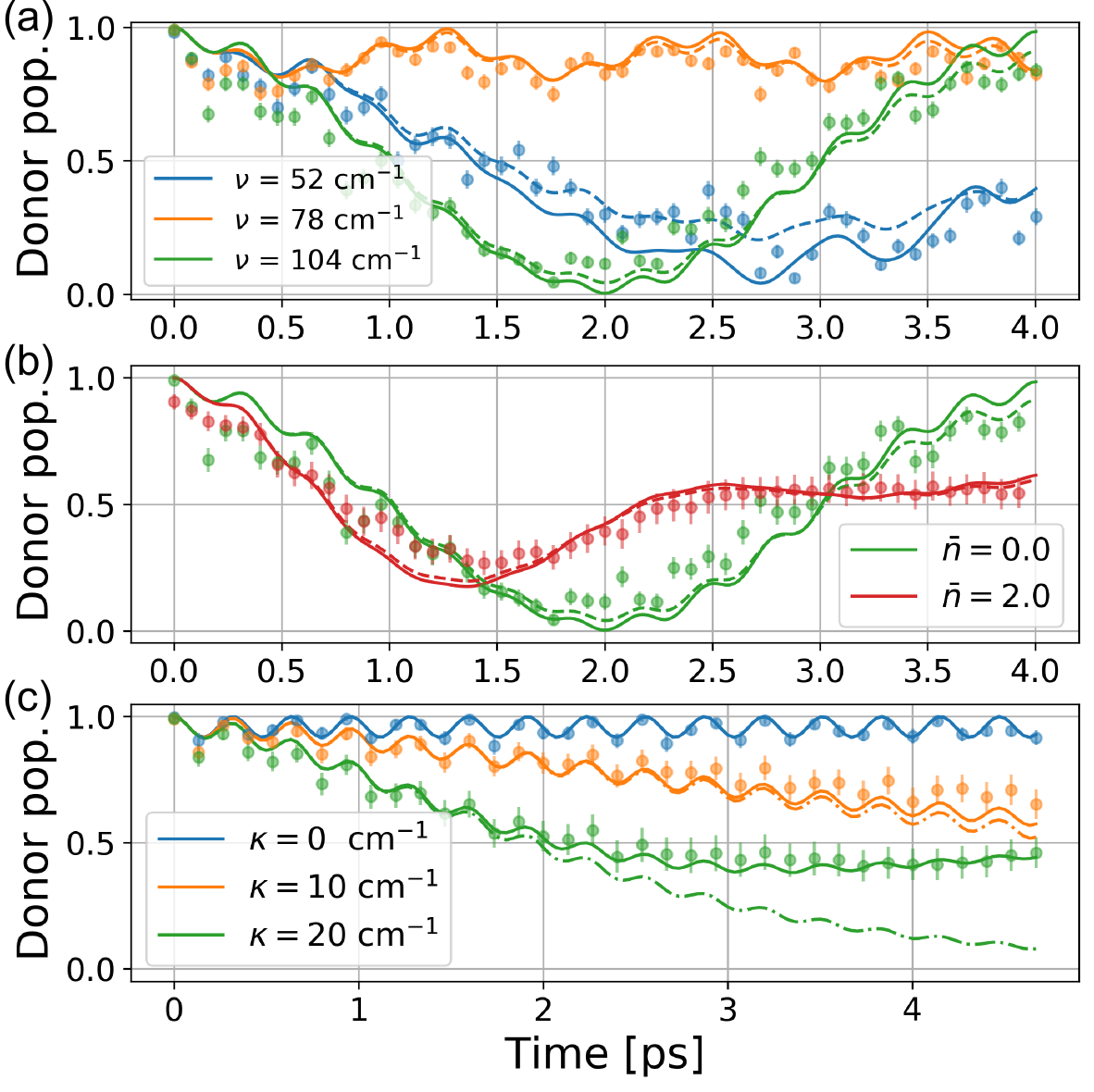}
\caption{\textbf{Simulations of vibration-assisted energy transfer (VAET).} We set $\epsilon=100\:\text{cm}^{-1}$ and $\Delta=30\:\text{cm}^{-1}$. \textbf{(a)} \textbf{(b)} Time evolution of the donor-state population for coherent VAET models with $\kappa=30\:\text{cm}^{-1}$. Solid and dashed curves depict theoretical predictions of the ideal target and the noise-aware model, respectively. Circles represent experimental data and error bars are standard deviation over 100 repetitions. In (a), bath temperature is fixed at \mbox{0 K} ($\bar{n}=0$) and $\nu$ is varied. In (b), $\nu$ is fixed to the resonant value $104\:\text{cm}^{-1}$ and two temperatures \mbox{0 K} and \mbox{300 K}, which correspond to $\bar{n}=0$ and 2.0, respectively, are simulated. \textbf{(c)} Time evolution of the donor-state population for dissipative VAET models with $\nu=104\:\text{cm}^{-1}$ and $\Gamma=10\:\text{cm}^{-1}$ at \mbox{0 K}. Solid and dot-dashed curves depict theoretical predictions of the dephased spin-oscillator model and the damped spin-oscillator model, respectively. Circles denote experimental data. For $\kappa=0\:\text{cm}^{-1}$, error bars indicate the standard deviation over 100 repetitions. For $\kappa=10$ and $20\:\text{cm}^{-1}$, error bars indicate the standard deviation over 20 trials, each trial performed with randomly drawn set of parameters and repeated 100 times.}\label{fig_VAET}
\end{figure}

Figure~\ref{fig_VAET}a shows the experimental data for simulating the coherent spin-single oscillator model $\hat{H}_{D}$ with a relatively large coupling strength (\mbox{$\kappa = 30 \: \text{cm}^{-1}$}). When $\nu$ satisfies the resonance condition of VAET (\mbox{$\nu = \sqrt{\Delta^2+\epsilon^2}/k \approx 104 \:\text{cm}^{-1}/k$} where $k$ is an integer), energy transfer occurs~\cite{Gorman18}. Otherwise, energy remains localized at the donor state. Figure \ref{fig_VAET}b shows the impact of initial temperature of the oscillator on the resonant VAET ($\nu=104 \: \text{cm}^{-1}$). We create baths of two temperatures \mbox{0 K} and \mbox{300 K} corresponding to initial average phonon number $\bar{n} = 0$ and $2.0$, respectively~\cite{supp}. The energy transfer is faster for higher $\bar{n}$ in the beginning as more vibrational quanta assist the energy transfer, but it also leads to a rapid decay of oscillations over time as the coherence is lost. 

In Fig.~\ref{fig_VAET}c, we simulate the dephased spin-oscillator model with $\Gamma=10\:\text{cm}^{-1}$ and various values of $\kappa$, to study the impact of dissipative coupling. We note that in the weak-$\kappa$ regime (or short evolution time), the results agree with the theoretical predictions of the \textit{damped} spin-oscillator model, which is equivalent to those of the spin-boson model with \mbox{$J(\omega) = J_{\rm Lo}(\omega)$}~\cite{Lemmer18, Tamascelli18}; however, as $\kappa$ is increased, the dephased and damped models' dynamics deviate. This is expected because unlike the damped model, the dephased model does not extract energy from the spin-oscillator system, so the donor population will not decay towards zero even after a long time evolution when coupled to a single-oscillator bath. This example demonstrates that our method provides the ability to simulate a wider range of dissipative channels in the study of open system dynamics.

\section*{Discussions and outlook}\label{disscusion}

%Simulations of large-scale spin-boson models with strong coupling in the presence of dissipation is crucial yet challenging problem in physical~\cite{Caldeira83, Weiss87}, chemical~\cite{Garg85, Gilmore05}, and biological~\cite{Huelga13, Jang18} systems. 
While classical simulation approaches to open dissipative systems continue to improve with advancements in numerical methods, they encounter fundamental challenges when dealing with large quantum systems that span exponentially large Hilbert space~\cite{MacDonell21, Kang24}. Take, for example, the Dissipation-Assisted Matrix Product Factorization (DAMPF) method, a state-of-the-art method for simulating open quantum systems~\cite{Somoza19}. For simulating a one-dimensional chain of $N_s$ electronic states each coupled to a bath of $N_b$ damped bosonic modes, the computational complexity of DAMPF scales as $\mathcal{O}(N_s^5N_bD_b^2\chi^3)$, where $D_b$ is the number of energy levels that represent each bosonic mode and $\chi$ is the bond dimension in the matrix product operator representation of the system-bath density matrix. A critical issue is the rapid, potentially exponential increase in $\chi$ with respect to the system-bath coupling strength, for a fixed target accuracy in the population dynamics. For these types of problems, trapped-ion simulators may offer a more scalable alternative, as the number of operations scales only as $\mathcal{O}(N_s N_b)$ and the total duration of SDK operations grows linearly with the system-bath coupling strength~\cite{Kang24}. The current state-of-the-art trapped-ion systems offer several dozens of highly coherent spins and bosonic modes that are very densely coupled with each other, the complete quantum description of which is totally intractable with classical computers of today~\cite{CetinaPRXQuantum2022,chen2023benchmarking,decross2024}. Developing various control methods over these quantum systems to study important physical, chemical, biological and material systems of critical interest will open up a new computational framework in science.

The methods developed in this work utilize fully programmable control methods to simulate dissipative processes. %This approach can readily be applied to existing simulations of chemical processes to include coupling to environment~\cite{Whitlow23,Valahu23,Sun23}. 
The dephased oscillator method can be used to realize non-Gaussian bath models~\cite{Barthel22}, relevant for studying physical processes such as phase transitions in strongly-coupled spin-boson models~\cite{Leggett87,Wang08,Duan17,Strathearn18,Nalbach13} and role of bosonic dephasing in molecular energy transfer~\cite{stock1990theory, banin1994impulsive}. The capability to measure the probability distribution of the motional modes~\cite{Fluehmann2020motion,Jia22,Valahu23,Whitlow23} may further reveal the effects of intricate vibronic interactions in biological and chemical reaction dynamics. High accuracy simulations will require development of efficient characterization methods~\cite{Kang23-E} and novel control protocols in the trapped-ion system~\cite{supp, lobser2023jaqalpaw} to implement various interaction terms in the target Hamiltonian. Combined with damped spin-boson models that can be implemented using sympathetic cooling~\cite{Lemmer18, Schlawin21}, the long-chain trapped-ion systems can provide a versatile platform for simulating dynamics of complex open quantum systems. %Additionally, the vibrational modes in our experiments  reaction coordinates, and their coupling with electronic states (spin in the manuscript) is essential for energy transfer efficiency. 

During the preparation of our manuscript we became aware of a complementary work~\cite{So24} which simulates the electron-transfer model using sympathetic cooling on the trapped ions' motional mode. \\

%\begin{acknowledgments}
{\bf Acknowledgments:} The authors thank Yichao Yu for his valuable discussions about the phase tracking method, Daniel Lobser for his valuable suggestion in controlling the RFSoC control software, and Thomas Barthel for his course on open quantum systems that inspired this work.
{\bf Funding:} This research is funded by the Office of the Director of National Intelligence - Intelligence Advanced Research Projects Activity, through the ARO contract W911NF-16-1-0082 (provided the experimental apparatus used in the demonstration), the DOE BES award DE-SC0019400 (theoretical modeling), and the NSF Quantum Leap Challenge Institute for Robust Quantum Simulation Grant No. OMA-2120757 (experimental concepts and validation). Support is also acknowledged from the U.S. Department of Energy, Office of Science, National Quantum Information Science Research Centers, Quantum Systems Accelerator (data analysis and validation).
{\bf Author contributions:} K.S., M.K., K.R.B. and J.K. conceived the idea and developed detailed methodology for experiments and modeling, K.S. and G.S. performed the experiments, M.K., H.N. and D.N.B performed modeling and simulations, K.S., M.K., K.R.B. and J.K. analyzed and validated the data, K.R.B. and J.K. supervised the project, K.S. and M.K. wrote the original draft, and K.S., M.K. and J.K. revised the manuscript. All authors discussed the results and reviewed the final manuscript.
{\bf Competing Interests:} J.K. and K.R.B are scientific advisors to IonQ, Inc., and J.K. is a shareholder of IonQ, Inc.
{\bf Data and materials availability:} All data needed to evaluate the conclusions in the paper are present in the paper and/or Supplementary Materials. We will provide raw data to anyone who might be interested.
%\end{acknowledgments}

\bibliography{maintxt}% common bib file
\end{document}

\title{Supplementary Material:\\Quantum Simulation of Spin-Boson Models with Structured Bath}

\author{Ke Sun}
\email{ke.sun@duke.edu}
\affiliation{Duke Quantum Center, Duke University, Durham, NC 27701, USA}
\affiliation{Department of Physics, Duke University, Durham, NC 27708, USA}

\author{Mingyu Kang}
\email{mingyu.kang@duke.edu}
\affiliation{Duke Quantum Center, Duke University, Durham, NC 27701, USA}
\affiliation{Department of Physics, Duke University, Durham, NC 27708, USA}

\author{Hanggai Nuomin}
\email{hanggai.nuomin@duke.edu}
\affiliation{Department of Chemistry, Duke University, Durham, NC 27708, USA}

\author{George Schwartz}
\affiliation{Duke Quantum Center, Duke University, Durham, NC 27701, USA}
\affiliation{Department of Physics, Duke University, Durham, NC 27708, USA}

\author{David N. Beratan}
\affiliation{Duke Quantum Center, Duke University, Durham, NC 27701, USA}
\affiliation{Department of Physics, Duke University, Durham, NC 27708, USA}
\affiliation{Department of Chemistry, Duke University, Durham, NC 27708, USA}
\affiliation{Department of Biochemistry, Duke University, Durham, NC 27710, USA}

\author{Kenneth R. Brown}
\affiliation{Duke Quantum Center, Duke University, Durham, NC 27701, USA}
\affiliation{Department of Physics, Duke University, Durham, NC 27708, USA}
\affiliation{Department of Chemistry, Duke University, Durham, NC 27708, USA}
\affiliation{Department of Electrical and Computer Engineering, Duke University, Durham, NC 27708, USA}

\author{Jungsang Kim}
\affiliation{Duke Quantum Center, Duke University, Durham, NC 27701, USA}
\affiliation{Department of Physics, Duke University, Durham, NC 27708, USA}
\affiliation{Department of Electrical and Computer Engineering, Duke University, Durham, NC 27708, USA}

\date{\today}

\maketitle

\section{Experimental Methods}

\subsection{Experimental implementation of spin-dependent kick operations}

\begin{figure*}[ht!]
\includegraphics[width=1.0\textwidth]{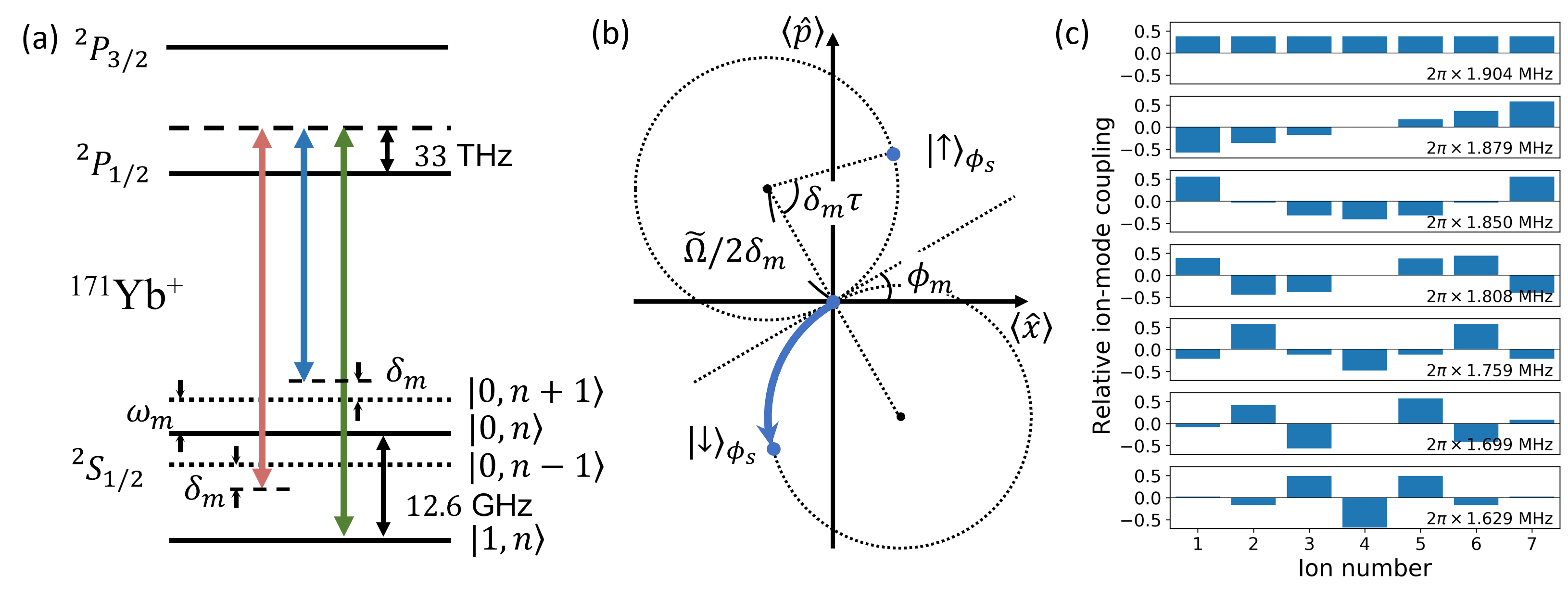}
\caption{\textbf{SDK operation and motional-mode structure. (a)} Simplified energy levels of the $^{171}$Yb$^+$ ion and the laser pulses used for implementing the on-resonance spin-dependent kick (SDK) operation. The global beam consists of two tones driving the blue sideband (BSB) and red sideband (RSB) transitions, accompanied by the individual beam (green arrow), respectively. \textbf{(b)} Visualization of the motional-state evolution in phase space when a SDK operation of duration $\tau$ is applied, where the qubit is in the eigenstate of the operator $\hat{\sigma}_{\phi_s} \equiv \hat{\sigma}_+ e^{i \phi_s} + \hat{\sigma}_- e^{-i \phi_s}$. Here, $\hat{\sigma}_+$ ($\hat{\sigma}_-$) is the raising (lowering) operator of the qubit, $\phi_s$ ($\phi_m$) is the spin (motion) phase of the laser beams, $\tilde{\Omega}$ is the Rabi frequency of the sideband transition, and $\delta_m$ is the motion detuning. \textbf{(c)} Normalized eigenvector components of each motional mode, which represent the relative ion-mode coupling strengths. Motional-mode frequencies are shown in the lower right corner of each panel.}\label{fig_exp}
\end{figure*}

The experiments are performed using a linear chain of $^{171}$Yb$^+$ ions. We use the hyperfine energy levels $\ket{0} \equiv \ket{F=1;m_F = 0}$ and $\ket{1} \equiv \ket{F=0;m_F = 0}$ in the $^2$S$_{1/2}$ manifold as the qubit states (spin states). The collective radial motional modes of the ion chain are used to represent the bosonic bath modes (see Figure~\ref{fig_exp}a). 

% The heating rate and decoherence time of the radial zig-zag mode in a 7-ion chain are $3.6(3)$ quanta/s and $5.2(7)$ ms, respectively. The other motional modes, except the center-of-mass mode, have similar noise parameters. See Ref.~\cite{Wang20} for the details of our experimental setup. 

Through the sequential application of Doppler cooling, EIT cooling, and sideband cooling, the zig-zag motional mode of the 7-ion chain can be cooled to average phonon number $\bar{n} = 0.036(16)$. Following this, a $370$ nm laser is utilized to pump the ions to their ground state $\ket{1}$. Transitions between the qubit states are performed using stimulated Raman transitions driven by a pair of laser beams~\cite{Wang20}. Acousto-optic modulators are employed to manipulate the frequency and phase of each laser beam, thereby enabling single-qubit gates; we apply a bit flip ($X$) gate to initialize the ion to the $\ket{0}$ state. The spin-oscillator coupling is simulated using the spin-dependent kick (SDK) operation induced by simultaneous blue- and red-sideband transitions as shown in Fig.~1 in the main text. Figure~\ref{fig_exp}b visualizes how the motional state evolves in phase space when a SDK operation of duration $\tau$, spin (motion) phase $\phi_s$ ($\phi_m$), motion detuning $\delta_m$, and sideband Rabi frequency $\tilde{\Omega}$ is applied~\cite{Lee05}. Subsequent to the final operation, the qubit-state population is accessed via the state-dependent detection technique.

The Lamb-Dicke parameters of the 7 motional modes range from 0.073 to 0.079 due to their different frequencies. The relative ion-mode coupling strengths (normalized eigenvector components) are shown in Fig.~\ref{fig_exp}c. We strategically use the center ion of a 7-ion chain, as it exhibits negligible coupling to the 2nd, 4th, and 6th modes. 
This arrangement maximizes the effective spacing between the motional-mode frequencies and therefore suppresses cross-mode coupling (driving modes that are not targeted by the SDK operation). 

% \begin{figure}[ht!]
% \includegraphics[width=0.45\textwidth]{Figures/Fig_mode.png}
% \caption{\textbf{Ion-Mode coupling strengths of a 7-ion chain.} Motional-mode frequencies are shown in the lower right corner of each panel. }\label{fig_mode}
% \end{figure}

\subsection{Trotterization}

The evolution of $\hat{H}_{D}$ in Eq.(2) of the main text is mapped to a sequence of single-qubit rotations and SDK operations via Trotterization. After conjugating $\hat{H}_{D}$ with a Hadamard gate on the qubit, the Hamiltonian can be described in the interaction picture as 
\begin{equation} \label{eq:HI}
    \hat{H}_I(t) = \frac{\epsilon}{2} \hat{\sigma}_{\Delta}(t) + \frac{\kappa}{2} \hat{\sigma}_{\Delta}(t) (\hat{b}e^{-i\nu t} + \hat{b}^\dagger e^{i \nu t}),
\end{equation}
where $\hat{\sigma}_{\Delta}(t) \equiv \hat{\sigma}_+ e^{i \Delta t} + \hat{\sigma}_- e^{-i \Delta t}$ and $\hat{\sigma}_+$ ($\hat{\sigma}_-$) is the raising (lowering) operator of the qubit. The evolution with respect to $\hat{H}_I(t)$ is discretized into $N$ time steps, each consisting of a single-qubit rotation and a SDK operation that simulate the first and second term, respectively. Specifically, for the SDK operation at the $j$-th time step, the spin (motion) phase of the laser beams~\cite{Lee05} is set as $\Delta t_j$ ($\nu t_j$), where $t_j \equiv (j-\frac{1}{2}) T / N \tau$ and $\tau$ is the duration of the operation for each time step. This is equivalent to setting the symmetric (anti-symmetric) detuning from the sideband transitions, referred to as the spin (motion) detuning, as $\Delta T / N \tau$ ($\nu T / N \tau$) at all time steps. This method can be straightforwardly extended to simulating models with multiple electronic states and vibrational modes~\cite{Sun23,Kang24}; however, the laser phase needs to be tuned to track the phase of each motional mode correctly (see Sec.~\ref{toolbox}). 

\subsection{Randomized operations}

Randomizing the control parameters is a key ingredient in our approach to simulating dissipation mechanisms in spin-boson models. By taking an average of many coherent evolutions, each involving a stochastically varying parameter, we realize both heating and dephasing as described by the Lindblad master equation~\cite{Chenu17}. Heating and dephasing are used to tune the initial temperature and spectral linewidth of the bath, respectively. 

First, the bosonic mode at finite temperature is created by modeling the mode's heating process as interacting with an infinite-temperature external bath for a finite duration. This heating is described by a pair of Lindblad operators \mbox{$\hat{L}_1 = \sqrt{\Gamma'} \hat{b}$} and \mbox{$\hat{L}_2 = \sqrt{\Gamma'} \hat{b}^\dagger$}, where $\Gamma'$ is the heating rate. We first prepare the motional mode at the ground state, and then apply $N'$ resonant ($\delta_m = 0$) SDK operations of duration \mbox{$\tau' = 4\Gamma'/\tilde{\Omega}^2$}, spin phase $\phi_s=0$, and motion phase $\phi_m$ randomly drawn from $[0, 2\pi)$ for each operation. These stochastic operations are applied on the initial qubit-mode composite state $\ket{+, n=0}$; note that $\ket{+}$ is an eigenstate of the spin operator with $\phi_s=0$, and thus the qubit is decoupled from these operations. When averaged over many trials of coherent evolutions, each executed with a distinct set of $N'$ random $\phi_m$ values, this procedure prepares the thermal state with an average phonon number $\bar{n} = N' \Gamma' \tau'$ (derivation provided in Sec.~\ref{sec:supp_randphase}). This can be intuitively understood as the thermal state represented as an ensemble of randomly displaced coherent states. 
For the experimental data in Fig.~2a and Fig.~5b of the main text, the average phonon number $\bar{n}$ of the initial thermal state is tuned by varying the duration $\tau'$ of SDK operations, where the number of steps $N'$ is fixed to 5. The errors in $\bar{n}$ due to finite number of steps are discussed in Sec.~\ref{finitesize}. 

\begin{figure}[ht!]
\includegraphics[width=1.0\textwidth]{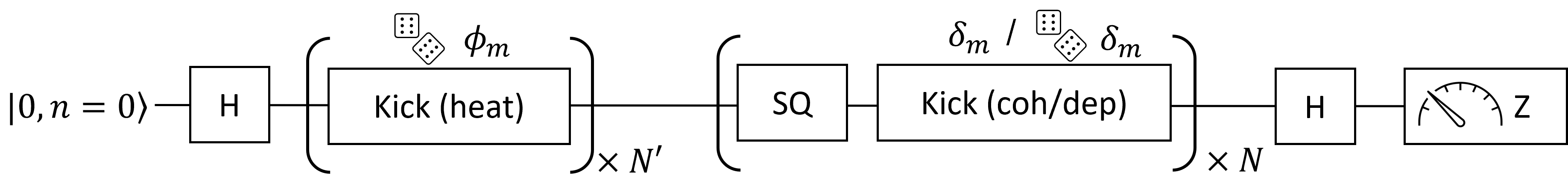}
\caption{\textbf{Instruction of operations for simulating the dephased spin-oscillator model.} `H' represents the Hadamard gate on the qubit. `Kick (heat)' represents the SDK operations with random $\phi_m$ that prepare the thermal state on average. `SQ' (`Kick (coh/dep)') represents the single-qubit rotations (SDK operations) that simulate the first (second) term in Eq.~\eqref{eq:HI}. Coherent or dephased spin-oscillator model is simulated by using a fixed or randomly drawn $\delta_m$ value at each step, respectively.}\label{fig_circuit}
\end{figure}

Next, dephasing during the time evolution, which is described by a single Lindblad operator \mbox{$\hat{L}_1 = \sqrt{\Gamma}\hat{b}^\dagger\hat{b}$}, is induced by giving random offsets to the motion detuning $\delta_m$ of the $N$ Trotterized SDK operations. Specifically, an offset to $\delta_m$ is assigned for each of the $N$ steps, with values drawn from a normal distribution characterized by a mean of zero and a standard deviation of $\sqrt{\Gamma T / N} / \tau$ (see Sec.~\ref{sec:supp_randdetuning}). The qubit population dynamics averaged over many trials of coherent evolutions, each performed with a set of $N$ random $\delta_m$ values, give the dynamics with respect to the dephased spin-oscillator model. The entire instruction of operations is shown in Fig.~\ref{fig_circuit}. 

In practice, the dephasing strength $\Gamma$ is varied by tuning the standard deviation of the random offsets to $\delta_m$ at each Trotterization step. This enables tuning the FWHM of each Lorentzian peak in the bath's spectral density. The errors in the simulated dephasing strength due to finite number of Trotterization steps are discussed in Sec.~\ref{finitesize}.

\section{Parameter mapping}
To illustrate how a trapped-ion simulator can simulate molecular energy transfer dynamics, we create a linear parameter mapping between the two systems. As a reference we define the unitless model with $\Delta = 1$. For the experiment in Fig. 3 of main text, the relationship between the unitless model and the molecular case is established by setting $\Delta = 500 \, \text{cm}^{-1}$~\cite{Chenu15}. The other frequency parameters are then scaled linearly with $\Delta$. The simulated evolution time $T_{\text{mol}}$ is determined from the unitless model's evolution time $T_{\text{ul}}$ by $\Delta_{\text{mol}} T_{\text{mol}} = \Delta_{\text{ul}} T_{\text{ul}}$, where the subscripts `mol' and `ul' represent molecular and unitless cases, respectively.

Similarly, the parameter $\kappa_{\text{ul},l} T_{\text{ul}} = \kappa_{\text{ion},l} T_{\text{ion},l}$ is used to determine the parameters for the trapped-ion simulator (`ion'), where $\kappa_{\text{ion},l}$ and $T_{\text{ion},l}$ denote the sideband Rabi frequency and the total operation time for driving the $l$-th mode, respectively. We choose $\kappa_{\text{ion},l}$ as reference for calculating other parameters because the sideband transition frequency is relatively challenging to adjust precisely. Subsequently, $\nu_{\text{ion},l}$ and $\Delta_{\text{ion},l}$ for each mode are determined by $\Delta_{\text{ul}} T_{\text{ul}} = \sum_l \Delta_{\text{ion},l} T_{\text{ion},l}$ and $\nu_{\text{ul},l} T_{\text{ul}} = \nu_{\text{ion},l} T_{\text{ion},l}$. The resulting parameter mapping relationships for the 3-mode case are shown in Table~\ref{table:Multi-mode}.

\begin{table}[h!]
\centering
\begin{tabular}{|c|c|c|c|c|c|c|c|c|c|c|c|c|}
\hline
\textbf{Model} & \multicolumn{3}{|c|}{\textbf{$\Delta$}} & \multicolumn{3}{|c|}{\textbf{$\kappa$}} & \multicolumn{3}{|c|}{\textbf{$\nu$}}& \multicolumn{3}{|c|}{Total time $(T)$} \\ \hline
Unitless & \multicolumn{3}{|c|}{1.0} & 0.1&0.1&0.1& 1.01&0.99&0.97 & \multicolumn{3}{|c|}{$12.8\times 2\pi$} \\ \hline
Molecular (cm$^{-1}$ or fs) &\multicolumn{3}{|c|}{500}  & 50&50&50 & 505&495&485 & \multicolumn{3}{|c|}{853.9} \\ \hline
Trapped ion ($2\pi\times $kHz or ms) & 7.11&4.78&3.98 & 2.13&1.43&1.19 & 21.54&14.20&11.59 & 0.60&0.89&1.07 \\ \hline
\end{tabular}
\caption{\textbf{Parameter mapping relations for the 3-mode experiment in Fig. 3 of main text.} For the trapped-ion case, the values for $\Delta$, $\kappa$, $\nu$, and $T$ correspond to the spin detuning, sideband Rabi frequency, motion detuning, and total operation time, respectively, used in the operations driving each motional mode.}
\label{table:Multi-mode}
\end{table}

We perform similar calculations for the VAET case, as shown in Table~\ref{table:VAET}. The presence of $\epsilon$ requires single-qubit operations during the Trotter steps. The average phonon number $\bar{n}$ in the molecular context can be determined by $\bar{n} = 1/(\exp(\hbar \nu / k_B \mathcal{T}) - 1)$, where $\mathcal{T} = 300 $K represents the room temperature. 

\begin{table}[h!]
\centering
\begin{tabular}{|c|c|c|c|c|c|c|}
\hline
\textbf{Model} & \textbf{$\epsilon$} & \textbf{$\Delta$} & \textbf{$\kappa$} & \textbf{$\nu$}& \multicolumn{2}{|c|}{Total time $(T)$} \\ \hline
Unitless & 1.0 & 0.3 & 0.3& 1.04 &  \multicolumn{2}{|c|}{$12\times 2\pi$} \\ \hline
Molecular (cm$^{-1}$ or ps) &100  & 30  & 30  & 104 & \multicolumn{2}{|c|}{4.003} \\ \hline
Trapped ion ($2\pi\times $kHz or ms) & 49.82 & 8 & 2.67 & 27.73 & $T_{\text{SQ}}=0.24$ & $T_{\text{kick}}=0.45$ \\ \hline
\end{tabular}
\caption{\textbf{Parameter mapping relations for simulating VAET in Fig. 5(b) of main text.} For the trapped-ion case, the value for $\epsilon$ represents the carrier Rabi frequency during the single-qubit rotations, and the values for $\Delta$, $\kappa$, $\nu$ correspond to the spin detuning, sideband Rabi frequency, and motion detuning, respectively, during the sideband-kick operations. Also, $T_{\text{SQ}}$ and $T_{\text{kick}}$ represent the total operation times for single-qubit rotations and sideband-kick operations, respectively.}
\label{table:VAET}
\end{table}

\section{Toolbox for achieving high-fidelity quantum simulation}\label{toolbox}
\subsection{Calibration}
The calibration of a trapped-ion quantum simulator is crucial for achieving accurate and reliable results. The simulator relies on precise control and measurement of individual ions, and any misalignment can lead to errors in simulation. Proper calibration ensures that the simulator operates within its specified parameters. Additionally, calibration is necessary for maintaining consistency in simulations and enabling reproducibility of experimentally obtained results. In addition to standard calibrated parameters such as Raman transition frequency and motional frequencies, more specific and refined calibration parameters are considered for this quantum simulation circuit. This section will predominantly focus on three calibration parameters that impact the fidelity of the simulation:
\begin{enumerate}
    \item Spin phase difference between single qubit (SQ) and spin-dependent kick (SDK) operations 
    \item Light shift difference between SQ and SDK operations
    \item Motional frequency re-calibration considering the light shift
\end{enumerate}

\subsubsection{Spin phase difference between SQ and SDK operations}
In the trapped-ion experimental setup, we execute the SQ operation ($\hat{\sigma}_{\alpha}$) using a carrier Raman transition and perform the SDK operations ($\hat{\sigma}_{\beta} (\hat{b} + \hat{b}^{\dagger})$) by simultaneously applying the blue- and red-sideband Raman transitions. Here, $\alpha$ in $\hat{\sigma}_{\alpha} \equiv \cos{\alpha} \hat{\sigma}_x + \sin{\alpha} \hat{\sigma}_y = e^{-i\alpha}\hat{\sigma}_+ + h.c.$ is the spin phase. In practice, the values of $\alpha$ and $\beta$ differ due to various factors:
\begin{enumerate}
    \item In the phase-sensitive geometry for Raman transitions, $\alpha - \beta = \pi/2$~\cite{wu2019quantum, jia2022determination} 
    \item The laser power and phase gradients at the ion position introduces extra spin phase
    \item Imperfect phase realization of the RFSoC system leads to a constant phase jump
\end{enumerate}

To explain the first two factors, we consider the Taylor expansion of the phase-sensitive Raman transition Hamiltonian, where a single ion is considered for simplicity. Under a specific rotating frame, the interaction Hamiltonian can be expressed as~\cite{jia2022quantum}
\begin{equation}
    \hat{H}_I = \frac{\Omega(x)}{2}e^{i(\Delta k x - \delta t -\Delta \phi)} \hat{\sigma}_+ + h.c.,
\end{equation}
where $x$ is the ion's displacement from the equilibrium position along the direction of wavevector difference $\Delta \vec{k}$ between the two Raman beams, $\Delta k \equiv |\Delta \vec{k}|$, $\Omega(x)$ is the Rabi frequency of the phase-sensitive Raman transition at displacement $x$, $\delta$ is the frequency detuning from the carrier transition, and $\Delta \phi$ is the phase difference between the two Raman beams. Approximating this expression to the first order with respect to $x$, we obtain
\begin{gather}
    \hat{H}_I \approx \left(\frac{\Omega(0)}{2}  + (\frac{\Omega'(0)}{2} + i\Delta k \frac{\Omega(0)}{2})x \right) e^{i(- \delta t -\Delta \phi)}\hat{\sigma}_+ + h.c.
\end{gather}
The position of ion can be quantized as 
\begin{equation}
    x = \sqrt{\frac{1}{2m\omega}} (\hat{b}e^{-i\omega t} + \hat{b}^{\dagger}e^{i\omega t})
\end{equation}
where $\omega$ denotes the motional frequency of this mode. Then, $\Delta k x = \eta (\hat{b}e^{-i\omega t} + \hat{b}^{\dagger}e^{i\omega t})$ where $\eta\equiv \Delta k \sqrt{\frac{1}{2m\omega}} $ represents the Lamb-Dicke parameter. Consequently, using the rotating-wave approximation (RWA), we can drive the on-resonance ($\delta=0$) SQ operation described by the Hamiltonian
\begin{align}
    \hat{H}_{SQ}  & =  \frac{\Omega(0)}{2} e^{-i\Delta \phi}\hat{\sigma}_+ + h.c.
\end{align}
Thus, the spin phase of the SQ operation is given by $\alpha = \Delta \phi$. When applying the SDK operation, we have
\begin{align}
    \hat{H}_{SDK} &= \hat{H}_I(\delta = -\omega) + \hat{H}_I(\delta = \omega)\nonumber \\
    & = (\frac{\Omega(0)'}{2}\frac{\eta}{\Delta k} + i\frac{\eta\Omega(0)}{2})e^{-i\Delta\phi} (\hat{b} + \hat{b}^{\dagger})\hat{\sigma}_+  + h.c. \nonumber \\ 
    & = |A|(\hat{b} + \hat{b}^{\dagger}) (e^{i\gamma}e^{-i\Delta \phi} \hat{\sigma}_+   + h.c.),
\end{align}
where $A = \frac{\Omega(0)'}{2}\frac{\eta}{\Delta k} + i\frac{\eta\Omega(0)}{2}$ and $\gamma = \text{Arg}(A)$. When devoid of laser power and phase gradient considerations, i.e. $\Omega'(0) = 0$ and $\Omega(0)$ is real, the spin phase of the SDK operation is given by $\beta = \Delta \phi - \pi/2$, which agrees with $\alpha-\beta = \pi/2$ as stated above. However, with the inclusion of laser power gradient considerations, $\gamma$ takes a nonzero value and introduces an additional spin phase component. Additionally, when considering the phase gradient of the laser wavepacket, $\Omega(x)$ is not necessarily a real number, introducing extra spin phase difference between the SQ and SDK operations. 

\begin{figure*}[ht!]
\centering
\includegraphics[width=1.0\textwidth]{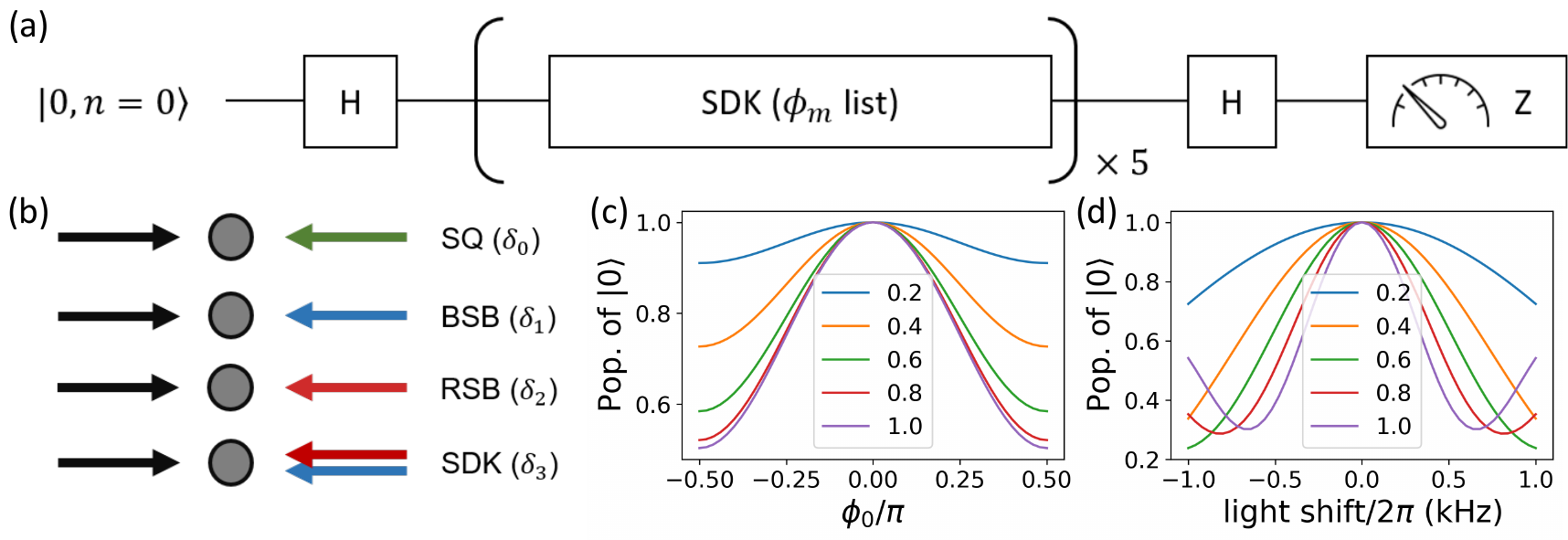}
\caption{\textbf{Calibration of light-shift differences.} \textbf{(a)} Illustration of pulse sequence used for calibrating the spin phase and light shift difference difference between the SQ and SDK operations. \textbf{(b)} Visual representation of the laser configurations for the SQ, blue/red sideband transition (BSB/RSB), and SDK operations. \textbf{(c)} \textbf{(d)} Expected probability of measuring the $\ket{0}$ state during the scanning of spin phase $\phi_0$ and light shift of the SDK operations, respectively. For the different curves, duration of the SDK operation for each step is varied, where the legend labels indicate the duration divided by the sideband $\pi$-time.}\label{fig_supp_cal_phase_ls}
\end{figure*}

In order to calibrate this spin phase difference between the SQ Hadamard gate and the SDK operation, the pulse sequence in Fig.~\ref{fig_supp_cal_phase_ls}a is employed, where the scanned spin phase of the SDK operation is denoted as $\phi_0$. The ion is initialized in the ground state $\ket{0, n = 0}$. Subsequent to the Hadamard gate, the ion's spin becomes $\ket{+} = \frac{1}{\sqrt{2}}(\ket{0}+\ket{1})$. Ideally, when $\phi_0$ is zero, the $\ket{+}$ state is the eigenstate of the SDK operation's spin operator and remains unchanged throughout. Following this, the second Hadamard gate reverts the spin state back to $\ket{0}$, resulting in a population measurement of 1 for observing $\ket{0}$. In contrast, if $\phi_0 \neq 0$, the spin state will undergo changes during the SDK operation, leading to a deviation from a population measurement of 1, as shown in Fig.~\ref{fig_supp_cal_phase_ls}c. Thus, the spin phase difference between the SQ and SDK operations can be calibrated by finding the $\phi_0$ value that gives the peak population of the $\ket{0}$ state. We note that the SDK operation is repeated five times where the list of motional phases is given by $[0, \pi, 0, \pi, 0]$ in order to prevent the ion from being excited to high motional states. Extending the SDK operation duration can amplify the accumulated errors and reduce the linewidth when scanning $\phi_0$, thereby producing a more accurate calibration outcome.

\subsubsection{Light shift difference between SQ and SDK operations}

When performing the SQ and SDK operations on a trapped-ion quantum simulator, different laser settings are employed. Specifically, during the application of the SQ gate, an individual beam and a global beam with the carrier frequency are directed at the ion, as depicted in Fig.~\ref{fig_supp_cal_phase_ls}b. Conversely, for SDK operations, the global beam carries two tones with respective BSB and RSB transition frequencies. As a result, the light shift, which depends on the laser intensity, frequency, and polarization, varies between these two cases.

In order to calibrate this light shift difference, we propose a similar calibration method as depicted in Fig.~\ref{fig_supp_cal_phase_ls}a. We iteratively perform the pule sequence while varying the light shift frequency of the SDK operation. When there is a non-zero light shift difference between SQ and SDK, an additional spin phase accumulates during the SDK operation, resulting in a mixed state at the end. Consequently, the population of $\ket{0}$ deviates from 1, as shown in Fig.~\ref{fig_supp_cal_phase_ls}d. Similarly to above, extending the SDK operation duration can enhance the accuracy of the calibration.

\subsubsection{Motional frequency re-calibration considering the light shift}

\begin{figure}[ht]
\centering
\includegraphics[width=0.7\textwidth]{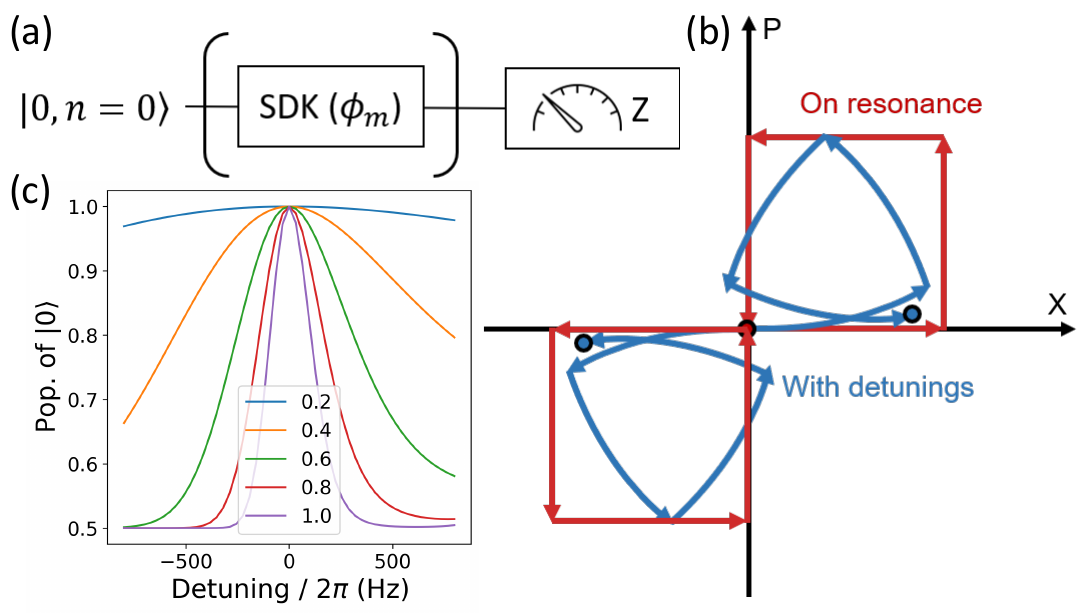}
\caption{\textbf{Motional frequency calibration.} \textbf{(a)} Illustration of pulse sequence used for calibrating the motional frequency. \textbf{(b)} Trajectory of ion in the phase space under this calibration scheme when the applied pulse is on resonance (red) or off resonance (blue). \textbf{(c)} Predictions of the measured population of the $\ket{0}$ state, where the motional detunings is scanned. The legend labels indicate the duration of each SDK operation divided by the sideband $\pi$-time.}\label{fig_supp_cal_motion}
\end{figure}

Common techniques for calibrating the BSB and RSB frequencies involve either direct measurement of the transition's spectrum or indirect determination of the detunings via Ramsey interferometric fringes. As depicted in Fig.~\ref{fig_supp_cal_phase_ls}b, the calibration process for BSB and RSB frequencies typically utilizes a laser beam at a constant frequency, while scanning the frequency of the other beam around the relevant frequency.  However, much like the situation described earlier, the light shifts experienced in these scenarios differ from those during the SDK operations, resulting in slight inaccuracies.

To address the potential inaccuracies in the calibration of motional frequencies, one straightforward approach is to employ the SDK operations for accurately determining the motional frequencies. This method is illustrated in Fig.~\ref{fig_supp_cal_motion}a, beginning with the ion initialized in the ground state for both its spin and motion ($\ket{n = 0}$). Subsequently, four SDK pulses are applied at motional phases of $0$, $\pi/2$, $\pi$, and $3\pi/2$. If the laser frequency perfectly matches the motional frequency, the ion's phase space trajectory will align with the red lines in Fig.~\ref{fig_supp_cal_motion}b, ultimately returning to the original pure state $\ket{0,n=0}$. However, if the laser frequency is detuned from the motional frequency, the ion will trace the blue trajectory, leading to a mixed state. When measuring the spin, the population of the $\ket{0}$ state will deviate from 1. 

By scanning the detuning of the laser and measuring the population of the $\ket{0}$ state, we can identify the maximum point, which corresponds to the motional frequency we aim to calibrate (Fig.~\ref{fig_supp_cal_motion}c). Moreover, increasing the SDK operation duration of each step reduces the linewidth of the scanned curve, improving the precision of the calibration. This method allows for precise calibration of the motional frequency, ensuring accurate and reliable simulations in the trapped ion quantum simulator.

\subsection{Phase tracking}

\begin{figure*}[ht!]
\centering
\includegraphics[width=1.0\textwidth]{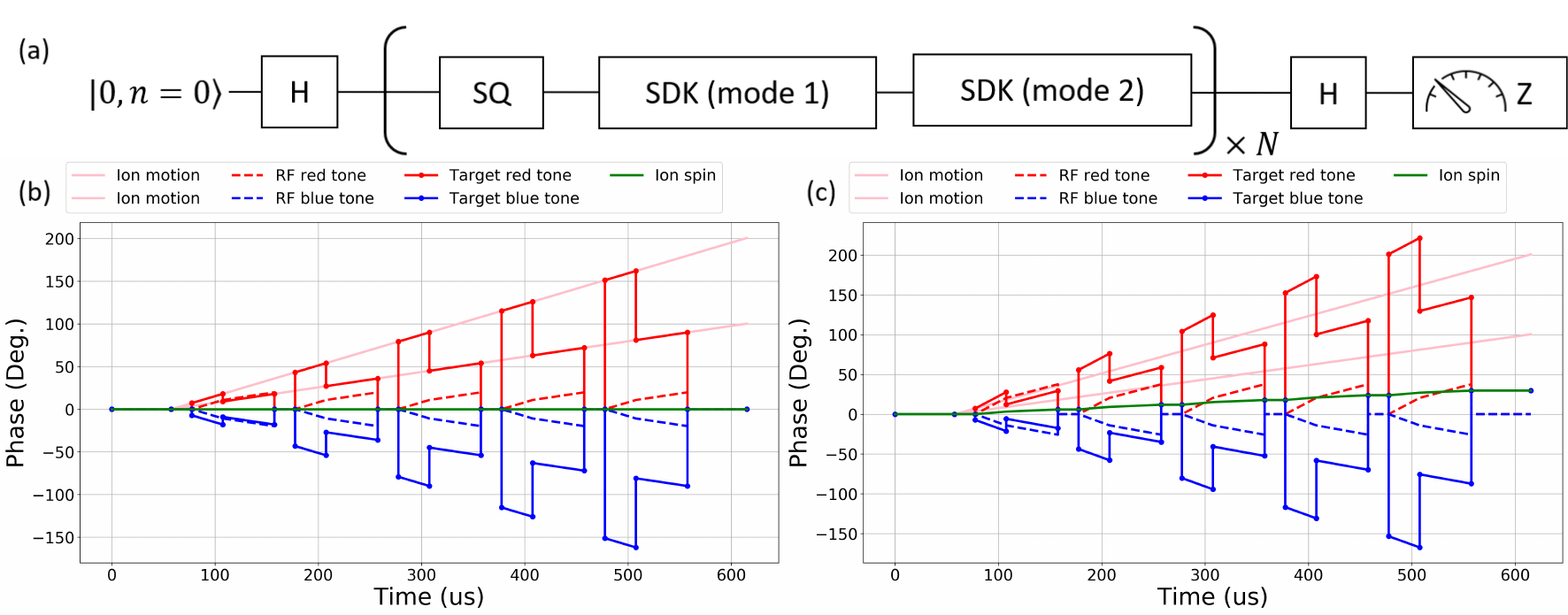}
\caption{\textbf{Demonstration of phase tracking.} \textbf{(a)} Pulse sequence used for showing the phases. \textbf{(b)} \textbf{(c)} Illustration of the evolution of different phases over time, including the spin and motion of the ion (ion spin/motion), the desired phases for implementing this pulse sequence (target red/blue tone), and the phases produced by the control software without applying additional phases (RF red/blue tone). Difference in light shifts between the SQ and SDK operations, as well as additional motional detunings, is assumed to be absent in (b) but present in (c).}\label{fig_supp_phase}
\end{figure*}

Once we have determined the frequencies and phases to be applied to the ion, it is essential to accurately implement the necessary control signals on the ion. Here we outline the procedures for controlling the laser phases to ensure synchronization with the ion phases. A simple pulse sequence shown in Fig.~\ref{fig_supp_phase}a is considered as an example that incorporates SQ rotations and SDK operations on two motional modes, where the number of Trotterization steps is set as $N=5$. It is important to note that the parameter values used in this example are chosen for clear visualization and thus are not at scale.

Figure~\ref{fig_supp_phase} displays the phases of the ion and the RFSoC control system. In this analysis, we utilize the frequency of the carrier transition ($\ket{0}$ to $\ket{1}$ transition) as the rotating frame. In practice, we set the repetition rate of the 355nm pulse laser to eliminate the light shift of the carrier transition. Thus, we assume the light shift of the carrier transition is zero. If not considering the light shift of the SDK operation, the spin phase of the ion remains constantly at zero in this frame, as indicated by the green line in Fig.~\ref{fig_supp_phase}b. Furthermore, regardless of the pulses applied to the ion, the motional modes oscillate at their respective motional frequencies. By setting the initial motional phase to zero at the start of the first SQ operation, the phase of the two motional modes of interest evolves as depicted by the two pink lines. To effectively monitor the ion's phase, we must align the phases of the lasers with the blue and red curves in Fig.~\ref{fig_supp_phase}b, which track the spin phase during the SQ operations and the blue and red motional phases during the SDK operations. Specifically, the target curves follow $\phi_{b,r} = \phi_s$ ($\phi_{b,r} = \phi_s \pm \phi_m$) for the SQ (SDK) operations. On the other hand, in the absence of additional phases applied to the RFSoC, the control system's phase follows the dashed curve if we sync the phase each time we apply an SQ operation. Consequently, we can determine the additional phase required to be added to the laser by subtracting the dashed curve from the solid curve.

We now take into account the difference in light shifts between the SQ and SDK operations. The green curve in Fig.~\ref{fig_supp_phase}c, which represents the ion spin's phase, exhibits a non-zero slope that is equivalent to the difference in light shifts, whenever a SDK operation is applied. Likewise, introducing additional detunings to the motional modes incurs the dashed curves to feature distinct slopes compared to the ion motion phase curves. Thus, the target phase curves need to be altered accordingly.

\section{Damped vs. dephased spin-oscillator model}

Two key ideas behind our experiment are (i) damped and dephased spin-oscillator models are described by the same 2-point correlation function (which determines the spectral density), and (ii) 2-point correlation function of the bath determines the spin dynamics up to leading order in the spin-bath coupling strength~\cite{OQS}. This enables our method of randomizing the control parameters to simulate the spin-boson model in the weak coupling regime.

\subsection{Correlation functions of spin-oscillator models}

We first introduce the 2-point correlation function of the bath. For simplicity, we consider a bath consisting of a single noisy oscillator; the correlation functions here can be straightforwardly generalized to a bath of multiple, or a continuum of, oscillators~\cite{Tamascelli18, Mascherpa20}. The spin-oscillator composite state $\rho$ follows the Lindblad master equation
\begin{equation} \label{eq:mastereqn}
    \dot{\rho} = -i [\hat{H}_{SO}, \rho] + \sum_j (\hat{L}_j \rho \hat{L}_j^\dagger - \frac{1}{2} \{ \hat{L}_j^\dagger \hat{L}_j, \rho \}),
\end{equation}
where $j$ is the index for different Lindblad operators $\hat{L}_j$ that act on the oscillator state only, and
\begin{equation}
    \hat{H}_{SO} = \hat{H}_S + \hat{A}_S \otimes \hat{F}_O + \hat{H}_O.
\end{equation}
Here, \mbox{$\hat{A}_S = \hat{\sigma}_Z$}/2, $\hat{F}_O = \kappa (\hat{b}+\hat{b}^\dagger)$, and $\hat{H}_O = \nu\hat{b}^\dagger\hat{b}$. We define the Lindblad generator $\mathcal{L}_O$ that describes the ``free evolution'' of the bath oscillator state $\rho_O$ as
\begin{equation}\label{eq:Lindbladgen}
    \mathcal{L}_O[\rho_O] = -i[\hat{H}_O, \rho_O] +  \sum_j (\hat{L}_j \rho_O \hat{L}_j^\dagger - \frac{1}{2} \{ \hat{L}_j^\dagger \hat{L}_j, \rho_O\}).
\end{equation}
In this case, the 2-point correlation function of the bath between times $t_0$ and $t_1$ ($t_0 > t_1$) is defined as~\cite{Tamascelli18, Mascherpa20}
\begin{align}
    C_O(t_0, t_1) &= {\rm Tr} \left\{ \hat{F}_O e^{\mathcal{L}_O (t_0 - t_1)} [\hat{F}_O e^{\mathcal{L}_O t_1}[\rho_O(0)]] \right\} \label{eq:COt0t1}\\
    &= {\rm Tr} \left\{ \hat{F}_O e^{\mathcal{L}_O t} [\hat{F}_O \rho_O(0)] \right\} = C_O(t), \label{eq:COt}
\end{align}
where $\rho_O(0)$ is the initial oscillator state and $t=t_0-t_1$. From Eq.~\eqref{eq:COt0t1} to \eqref{eq:COt}, we use the common assumption that the initial state is in equilibrium with dissipation, i.e., $\mathcal{L}_O[\rho_O(0)] = 0$. In the Heisenberg picture, Eq.~\eqref{eq:COt} can be written as
\begin{equation}\label{eq:COtHeisenberg}
    C_O(t) = {\rm Tr} \left\{ e^{\mathcal{L}_O^\dagger t}[\hat{F}_O] \hat{F}_O \rho_O(0) \right\},
\end{equation}
where $\mathcal{L}_O^\dagger$ is the \textit{adjoint} Lindblad generator, such that~\cite{OQS}
\begin{equation}\label{eq:Lindbladgenadj}
    \mathcal{L}_O^\dagger[\hat{F}_O] = i[\hat{H}_O, \hat{F}_O] +  \sum_j (\hat{L}_j^\dagger \hat{F}_O \hat{L}_j - \frac{1}{2} \{ \hat{L}_j^\dagger \hat{L}_j, \hat{F}_O\}).    
\end{equation}

The higher-order correlation functions can be defined as well. For example, the 4-point correlation function of the bath in the noisy spin-oscillator model can be written as
\begin{align} \label{eq:4ptcorrfn}
C^{(4)}_O(t_0, t_1, t_2, t_3) &= {\rm Tr} \left\{ 
\hat{F}_O e^{\mathcal{L}_O (t_0-t_1)}[ 
\hat{F}_O e^{\mathcal{L}_O (t_1-t_2)}[
\hat{F}_O e^{\mathcal{L}_O (t_2-t_3)}[
\hat{F}_O\rho_O(0)]]] \right\} \\
&= {\rm Tr} \left\{ 
e^{\mathcal{L}_O^\dagger (t_0 - t_1)}[
e^{\mathcal{L}_O^\dagger (t_1 - t_2)}[
e^{\mathcal{L}_O^\dagger (t_2 - t_3)}[\hat{F}_O]
\hat{F}_O]\hat{F}_O] \hat{F}_O \rho_O(0)\right\}
\end{align}
where $t_0 > t_1 > t_2 > t_3$ and we again assume the initial-state equilibrium $\mathcal{L}_O[\rho_O(0)] = 0$. 

Now we compare the correlation functions of the bath for the damped and dephased spin-oscillator models. First, damping is described by a pair of Lindblad operators \mbox{$\hat{L}_1 = \sqrt{\Gamma (\bar{n}+1)} \hat{b}$} and \mbox{$\hat{L}_2 = \sqrt{\Gamma \bar{n}} \hat{b}^\dagger$}, where the bath oscillator's initial state $\rho_O(0)$ is given by the thermal state with average phonon number $\bar{n}$. The Lindblad generator $\mathcal{L}_O^{\rm (damp)}$, obtained by substituting these $\hat{L}_j$'s into Eq.~\eqref{eq:Lindbladgen}, satisfies $\mathcal{L}_O^{\rm (damp)}[\rho_O(0)] = 0$ such that Eq.~\eqref{eq:COt} is valid. 

Next, dephasing is described by a single Lindblad operator $\hat{L}_1 = \sqrt{\Gamma} \hat{b}^\dagger \hat{b}$. The Lindblad generator $\mathcal{L}_O^{\rm (deph)}$, obtained by substituting this $\hat{L}_j$ into Eq.~\eqref{eq:Lindbladgen}, satisfies  $\mathcal{L}_O^{\rm (deph)}[\rho_O(0)] = 0$ for any diagonal state $\rho_O(0)$ such that Eq.~\eqref{eq:COt} is valid. 

Now we substitute $\hat{F}_O = \kappa(\hat{b}+\hat{b}^\dagger)$, which represents the linear spin-oscillator coupling, into Eq.~\eqref{eq:Lindbladgenadj}. Interestingly, the Lindblad generators of the damped and dephased oscillators give the same result:
\begin{equation}
    \mathcal{L}_O^{\rm (damp)\dagger}[\hat{F}_O] = \mathcal{L}_O^{\rm (deph)\dagger}[\hat{F}_O] = \kappa \left( (-i\nu - \frac{\Gamma}{2}) \hat{b} + (i \nu - \frac{\Gamma}{2}) \hat{b}^\dagger \right).
\end{equation}
Plugging this into Eq.~\eqref{eq:COtHeisenberg}, we obtain the bath's 2-point correlation function of both the damped and dephased spin-oscillator models, given by
\begin{equation} \label{eq:COtdampdeph}
    C_O(t) = \kappa^2 e^{-\Gamma t/2} \left(\coth(\frac{\beta\nu}{2}) \cos \nu t - i \sin \nu t \right),
\end{equation}
where $\beta = \log(1+1/\bar{n})/\nu$ is the inverse of the temperature. 

\begin{figure*}[ht!]
\centering
\includegraphics[width=0.9\textwidth]{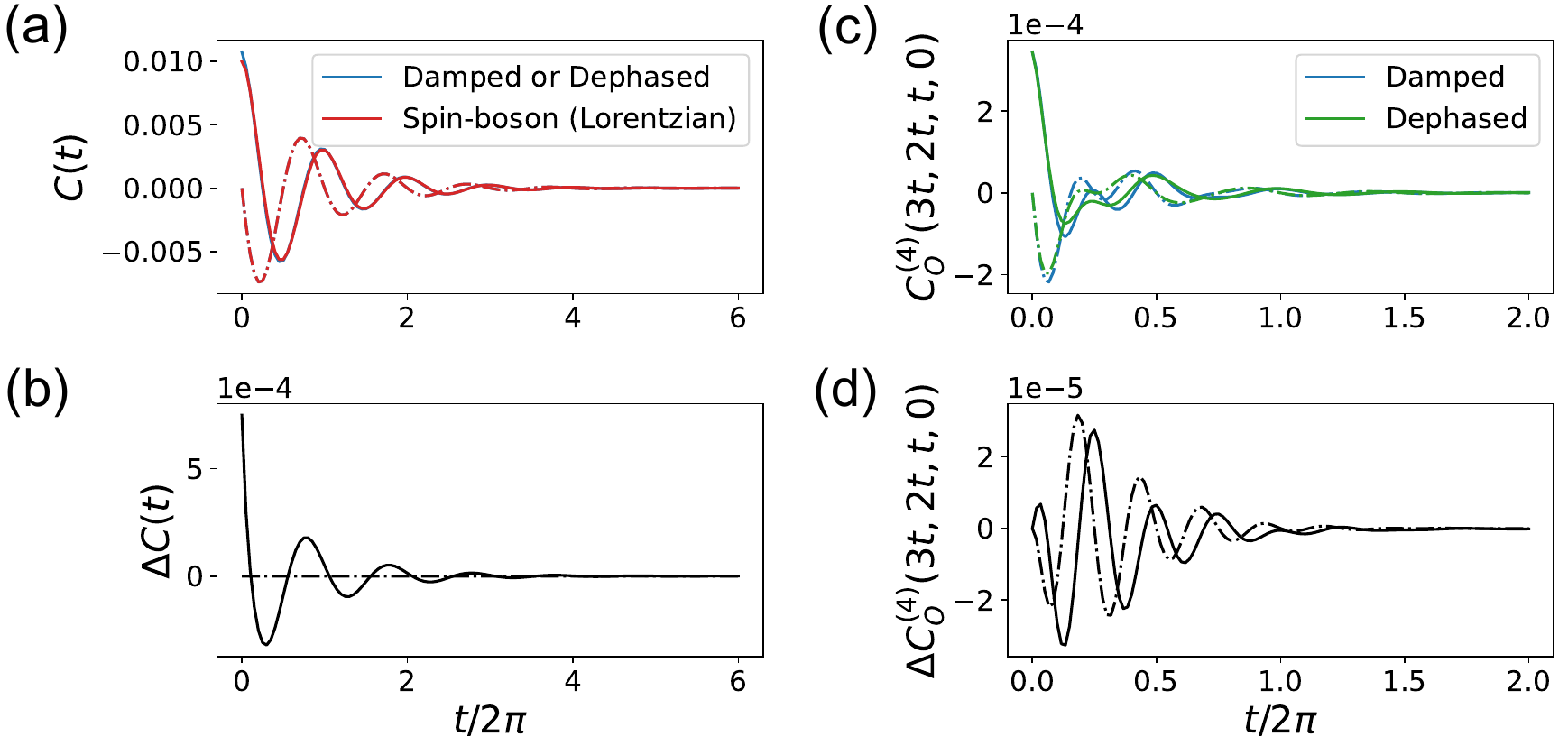}
\caption{\textbf{Correlation functions.} In all panels, $\nu=1$, $\Gamma=0.4$, $\kappa=0.1$, and $\beta = 3.36$ ($\bar{n} = 0.036$). Solid (dot-dashed) curve is the real (imaginary) part. \textbf{(a)} 2-point correlation functions of the damped or dephased spin-oscillator model [see Eq.~\eqref{eq:COtdampdeph}] and the spin-boson model with Lorentzian spectral density $J(\omega) = J_{\rm Lo}(\omega)$ [see Eq.~\eqref{eq:CtJomega}], which match reasonably well. \textbf{(b)} Difference between the 2-point correlation functions in (a). \textbf{(c)} Examples of 4-point correlation functions of the damped and dephased spin-oscillator models, numerically calculated from Eq.~\eqref{eq:4ptcorrfn}. The results closely match Eqs.~\eqref{eq:C4damp} and \eqref{eq:C4deph} obtained for $\bar{n}=0$. \textbf{(d)} Difference between the 4-point correlation functions in (c).}\label{fig_supp_corrfn}
\end{figure*}

Note that the 2-point correlation function $C(t)$ and the spectral density $J(\omega)$ are closely related descriptions of the bath. In particular, for the spin-boson model, in which the bath is a continuum of oscillators, the two quantities are related by 
\begin{equation}\label{eq:CtJomega}
C(t) = \int_0^\infty d\omega \frac{J(\omega)}{\pi} \left( \coth(\frac{\beta \omega}{2}) \cos \omega t - i \sin \omega t \right).  
\end{equation}
For the damped spin-oscillator model, this relation is approximately true for the Lorentzian spectral density $J(\omega) = J_{\rm Lo}(\omega)$ in a reasonable regime of parameters (\mbox{$\Gamma < \nu/2$}, \mbox{$\beta \ll 2\pi (\Gamma/2)^{-1}$})~\cite{Lemmer18}. See Fig.~\ref{fig_supp_corrfn}a,b for the numerical agreement. Thus, in this paper, we conveniently use Eq.~\eqref{eq:CtJomega} to describe the spectral density of both the damped and dephased spin-oscillator models. As the 2-point correlation functions of the two models are the same as in Eq.~\eqref{eq:COtdampdeph}, the same Lorentzian spectral density $J_{\rm Lo}(\omega)$ is assigned to the dephased spin-oscillator model. We leave a rigorous derivation of the dephased spin-oscillator model's spectral density, similarly to the work on the damped model~\cite{Grabert84}, to future research. 

While the 2-point correlation functions of the damped and dephased spin-oscillator models match, the 4-point correlation functions do not. This can be easily understood from the fact that $\mathcal{L}_O^{\rm (damp)\dagger}[\hat{b}] = \mathcal{L}_O^{\rm (deph)\dagger}[\hat{b}]$ and $\mathcal{L}_O^{\rm (damp)\dagger}[\hat{b}^\dagger] = \mathcal{L}_O^{\rm (deph)\dagger}[\hat{b}^\dagger]$ but $\mathcal{L}_O^{\rm (damp)\dagger}[\hat{b}^\dagger \hat{b}] \neq \mathcal{L}_O^{\rm (deph)\dagger}[\hat{b}^\dagger \hat{b}]$. For simplicity, we only show the 4-point correlation functions for the zero-temperature ($\bar{n}=0$) case:
\begin{align}
{\rm Damped: } \quad & C^{(4)}_O(t_0, t_1, t_2, t_3) =  \kappa^4 \left(e^{(-\Gamma/2 - i \nu) (t_0-t_1+t_2-t_3)} + 2 e^{(-\Gamma/2 - i \nu) (t_0+t_1-t_2-t_3)} \right), \label{eq:C4damp}\\
{\rm Dephased: } \quad & C^{(4)}_O(t_0, t_1, t_2, t_3) = \kappa^4 \left( e^{(-\Gamma/2 - i \nu) (t_0-t_1+t_2-t_3)} + 2 e^{-\Gamma(t_1-t_2)} e^{(-\Gamma/2 - i \nu) (t_0+t_1-t_2-t_3)} \right). \label{eq:C4deph}
\end{align}
Examples of 4-point correlation functions are shown in Fig.~\ref{fig_supp_corrfn}c,d. Higher-order (6-point, 8-point, ...) correlation functions are also different between the two models. 

Note that for the damped case, the 4-point correlation function can be expressed only using the 2-point correlation functions as 
\begin{equation}
{\rm Damped: } \quad C^{(4)}_O(t_0, t_1, t_2, t_3) = C_O(t_0, t_1) C_O(t_2, t_3) + C_O(t_0,t_2) C_O(t_1,t_3) + C_O(t_0,t_3)C_O(t_1,t_2),  
\end{equation}
which is a signature that the damping is Gaussian~\cite{Linowski22}, i.e., the oscillator state is Gaussian at all times given a Gaussian initial state~\cite{Tamascelli18}. This can be understood from the fact that each term of $\mathcal{L}_O$ in Eq.~\eqref{eq:Lindbladgen} contains up to only two $\hat{b}$ or $\hat{b}^\dagger$. However, dephasing is not Gaussian, as terms in $\mathcal{L}_O$ contain at most four $\hat{b}$ or $\hat{b}^\dagger$. Thus, the 4-point correlation function cannot be expressed only using the 2-point correlation functions. 

\subsection{Spin dynamics and correlation functions}

Following Ref.~\cite{OQS}, the spin dynamics of the models here can be described in terms of the bath's correlation functions. We first transform the Hamiltonian $\hat{H}_{SO}$ into the interaction picture as
\begin{equation}
    \hat{H}_{SO}'(t) = \hat{A}'_S(t) \otimes \hat{F}'_O(t), 
\end{equation}
where $\hat{A}'_S(t) := e^{i \hat{H}_S t} \hat{A}_S e^{-i \hat{H}_S t}$ and $\hat{F}'_O(t) := e^{\mathcal{L}_O^\dagger t} [\hat{F}_O]$. The spin-oscillator density matrix $\rho'(t)$ in the interaction picture evolves with respect to the generator $\mathcal{L}'(t)$ as 
\begin{equation}
    \dot{\rho}'(t) = \mathcal{L}'(t)[\rho'(t)] = -i[\hat{H}_{SO}'(t), \rho'(t)].
\end{equation} 

Note that interaction picture is typically defined for the case where a static Hamiltonian, rather than a Lindblad generator such as $\mathcal{L}_O$, is rotated out. Thus, for mathematical rigor, the noisy spin-oscillator model first needs to be embedded into a tripartite spin-oscillator-environment system undergoing a corresponding unitary evolution, as in Lemma 1 of Ref.~\cite{Tamascelli18}. Then, the interaction picture is obtained by rotating out the Hamiltonian of the oscillator-environment subsystem. 

To extract the spin dynamics, we use the time-convolutionless (TCL) projection operator method~\cite{Tokuyama76, OQS}. We introduce the projection operator $\mathcal{P}$, defined as
\begin{equation}
    \mathcal{P} \rho'(t) = \rho_S'(t) \otimes \rho_O(0),
\end{equation}
where $\rho_S'(t) = {\rm Tr}_O \{\rho'(t)\}$, such that $\mathcal{P} \rho'(t)$ contains the information about the spin part of $\rho'(t)$. Then we obtain the TCL master equation for $\mathcal{P} \rho'(t)$ as  
\begin{equation} \label{eq:TCLmastereqn}
    \frac{\partial}{\partial t} \mathcal{P} \rho'(t) = \sum_{n=1}^\infty \mathcal{K}_n(t) \mathcal{P} \rho'(t),
\end{equation}
where $\mathcal{K}_n(t)$ is the $n$-th order TCL generator. The TCL generators up to fourth order are given by
\begin{align}
    \mathcal{K}_1(t) &= \mathcal{K}_3(t) = 0,\\
    \mathcal{K}_2(t) &= \int_0^t dt_1 \mathcal{P} \mathcal{L}'(t) \mathcal{L}'(t_1) \mathcal{P},\\
    \mathcal{K}_4(t) &= \int_0^t dt_1 \int_0^{t_1} dt_2 \int_0^{t_2} dt_3 \Big(
\mathcal{P} \mathcal{L}'(t) \mathcal{L}'(t_1) \mathcal{L}'(t_2) \mathcal{L}'(t_3) \mathcal{P}
- \mathcal{P} \mathcal{L}'(t) \mathcal{L}'(t_1) \mathcal{P} \mathcal{L}'(t_2) \mathcal{L}'(t_3) \mathcal{P} \nonumber \\
&\quad\quad\quad\quad\quad\quad\quad\quad\quad\quad\:\:
- \mathcal{P} \mathcal{L}'(t) \mathcal{L}'(t_2) \mathcal{P} \mathcal{L}'(t_1) \mathcal{L}'(t_3) \mathcal{P}
- \mathcal{P} \mathcal{L}'(t) \mathcal{L}'(t_3) \mathcal{P} \mathcal{L}'(t_1) \mathcal{L}'(t_2) \mathcal{P}
\Big). \label{eq:K4t}
\end{align}
In Eq.~\eqref{eq:K4t}, the first term of the integrand starts and ends with $\mathcal{P}$, while the other three terms have another $\mathcal{P}$ multiplied in the middle. The odd-order TCL generators are zero because they involve odd number of $\hat{b}$ or $\hat{b}^\dagger$ applied to the diagonal state $\rho_O(0)$, followed by $\mathcal{P}$ which traces out the oscillator. Also note that for even $n$, $\mathcal{K}_n(t)$ is proportional to $\kappa^n$, as $\hat{H}'_{SO}(t) \propto \kappa$. Thus, the TCL master equation in Eq.~\eqref{eq:TCLmastereqn} can be thought as a perturbation expansion with respect to the spin-oscillator coupling strength $\kappa$. 

We first write the second-order term of Eq.~\eqref{eq:TCLmastereqn} using the bath's 2-point correlation function $C_O(t, t_1)$. Here we define $\hat{0} \equiv \hat{A}'_S(t)$ and $\hat{1} \equiv \hat{A}'_S(t_1)$, using the notation of Ref.~\cite{OQS}. By straightforward algebra, 
\begin{align}\label{eq:K2}
    \mathcal{K}_2(t) \rho'_S(t) \otimes \rho_O(0) 
    &= -\int_0^t dt_1 {\rm Tr}_O \left\{[\hat{H}'_{SO}(t), [\hat{H}'_{SO}(t_1),\rho'_S(t) \otimes \rho_O(0) ]] \right\} \otimes \rho_O(0) \nonumber\\
    &= -\int_0^t dt_1 {\rm Tr}_O \Big\{ 
    \hat{0}\hat{1}\rho'_S(t) \otimes \hat{F}'_O(t) \hat{F}'_O(t_1) \rho_O(0) 
    + \rho'_S(t)\hat{1}\hat{0} \otimes \rho_O(0) \hat{F}'_O(t_1)\hat{F}'_O(t) \nonumber\\
    &\quad\quad\quad\quad\quad\quad\:
    - \hat{0}\rho'_S(t)\hat{1} \otimes \hat{F}'_O(t)\rho_O(0)\hat{F}'_O(t_1)
    - \hat{1}\rho'_S(t)\hat{0} \otimes \hat{F}'_O(t_1)\rho_O(0)\hat{F}'_O(t)
    \Big\}\otimes \rho_O(0) \nonumber\\
    &= -\int_0^t dt_1 \Big( 
    C_O(t, t_1)(\hat{0}\hat{1}\rho'_S(t) - \hat{1}\rho'_S(t)\hat{0}) 
    - C_O^*(t, t_1) (\hat{0}\rho'_S(t)\hat{1} - \rho'_S(t)\hat{1}\hat{0})
    \Big) \otimes \rho_O(0),
\end{align} 
where $^*$ denotes complex conjugate and we use Eq.~\eqref{eq:COt0t1} to go from second to third line. Thus, separating out $\rho_O(0)$ from both sides, this gives the $\mathcal{O}(\kappa^2)$ term of the time evolution of the spin.

The fourth-order term of Eq.~\eqref{eq:TCLmastereqn} can be written in terms of the bath's 4-point correlation function in Eq.~\eqref{eq:4ptcorrfn}. Using the notation $\hat{2} \equiv \hat{A}'_S(t_2)$ and $\hat{3} \equiv \hat{A}'_S(t_3)$, similar algebra yields
\begin{align}\label{eq:K4}
    \mathcal{K}_4(t) \rho'_S(t) &\otimes \rho_O(0) 
    = \int_0^t dt_1 \int_0^{t_1} dt_2 \int_0^{t_2} dt_3 \nonumber \\
    &\quad \Big( 
    C^{(4)}_O(t, t_1, t_2, t_3) (\hat{0}\hat{1}\hat{2}\hat{3}\rho'_S(t) - \hat{1}\hat{2}\hat{3}\rho'_S(t)\hat{0}) 
    - C^{(4)*}_O(t, t_1, t_2, t_3) (\hat{0}\rho'_S(t)\hat{3}\hat{2}\hat{1} - \rho'_S(t)\hat{3}\hat{2}\hat{1}\hat{0}) \nonumber \\
    & - C^{(4)}_O(t_2, t, t_1, t_3) (\hat{0}\hat{1}\hat{3}\rho'_S(t)\hat{2} - \hat{1}\hat{3}\rho'_S(t)\hat{2}\hat{0})
    + C^{(4)*}_O(t_2, t, t_1, t_3) (\hat{0}\hat{2}\rho'_S(t)\hat{3}\hat{1} - \hat{2}\rho'_S(t)\hat{3}\hat{1}\hat{0}) \nonumber \\
    & - C^{(4)}_O(t_1, t, t_2, t_3) (\hat{0}\hat{2}\hat{3}\rho'_S(t)\hat{1} - \hat{2}\hat{3}\rho'_S(t)\hat{1}\hat{0})
    + C^{(4)*}_O(t_1, t, t_2, t_3) (\hat{0}\hat{1}\rho'_S(t)\hat{3}\hat{2} - \hat{1}\rho'_S(t)\hat{3}\hat{2}\hat{0}) \nonumber \\
    & + C^{(4)}_O(t_2, t_1, t, t_3) (\hat{0}\hat{3}\rho'_S(t)\hat{2}\hat{1} - \hat{3}\rho'_S(t)\hat{2}\hat{1}\hat{0}) 
    - C^{(4)*}_O(t_2, t_1, t, t_3) (\hat{0}\hat{1}\hat{2}\rho'_S(t)\hat{3} - \hat{1}\hat{2}\rho'_S(t)\hat{3}\hat{0})    
    \Big) \otimes \rho_O(0).
\end{align}
Again, separating out $\rho_O(0)$ from both sides, this gives the $\mathcal{O}(\kappa^4)$ term of the spin state's evolution. 

Now we return to the damped and dephased spin-oscillator models. In the previous subsection, we showed that the 2-point correlation functions match between the two models, but the 4-point correlation functions do not. Thus, combined with Eqs.~\eqref{eq:K2} and \eqref{eq:K4}, this leads to the conclusion that the spin-state dynamics of the two models agree up to $\mathcal{O}(\kappa^2)$, but not to $\mathcal{O}(\kappa^4)$. 

Let $\hat{O}_S(t)$ be a spin observable of interest in the interaction picture. The difference $|\Delta \langle \hat{O}_S(t) \rangle|$ in the expectation value of $\hat{O}_S(t)$ between the damped and dephased spin-oscillator models can be upper bounded using the difference $\Delta C_O^{(4)}(t_0, t_1, t_2, t_3)$ between the 4-point correlation functions in Eqs.~\eqref{eq:C4damp} and \eqref{eq:C4deph}. Let $||\cdot||$ denote the max absolute eigenvalue of the argument. By integrating Eq.~\eqref{eq:K4} over time and then applying inequalities such as $|{\rm Tr}\{\hat{O}_S(t)\hat{0}\hat{1}\hat{2}\hat{3}\rho'_S(t)\}| \leq ||\hat{O}_S|| \times ||\hat{A}_S||^4 = ||\hat{O}_S||/16$, we obtain
\begin{align} \label{eq:C4bound}
    |\Delta \langle \hat{O}_S(t) \rangle| \leq \frac{||\hat{O}_S||}{4}
    \int_0^t dt_0 \int_0^{t_0} dt_1 \int_0^{t_1} dt_2 \int_0^{t_2} dt_3 
    \Big(&|\Delta C^{(4)}_B(t_0, t_1, t_2, t_3)| + |\Delta C^{(4)}_B(t_2, t_0, t_1, t_3)| \nonumber \\
    + &|\Delta C^{(4)}_B(t_1, t_0, t_2, t_3)| + |\Delta C^{(4)}_B(t_2, t_1, t_0, t_3)|
    \Big) + \mathcal{O}(\kappa^6).
\end{align}
Note that this upper bound is in similar form to the second-leading order term in Eq.~(6) of Ref.~\cite{Mascherpa17} (when the exponential is expanded into a Taylor series), which is derived for two spin-boson models with different 2-point correlation functions. For the case of spin-boson models, it is sufficient to consider the difference in only the 2-point correlation functions, as the spin-boson model's bath $\hat{H}_B = \int_0^\infty d\omega \hat{a}^\dagger(\omega) \hat{a}(\omega)$ is Gaussian. However, the \textit{dephased} spin-oscillator model is non-Gaussian; thus, it is necessary to consider the higher-order correlation functions for obtaining such bounds. In practice, for the difference between the damped and dephased spin-oscillator models, the bound in Eq.~\eqref{eq:C4bound} grows much faster with $t$ than the actual $|\Delta \langle \hat{O}_S(t) \rangle|$ obtained from solving the Lindblad master equations.

\section{Randomized operations}\label{randomization}

In this section, we explain our method of using randomized control parameters on the quantum harmonic oscillator to (i) prepare the thermal state and (ii) simulate the dephasing. We use the fact that heating and dephasing can be described as an average of many evolutions of closed systems, where the parameter of each closed system's Hamiltonian is randomly drawn~\cite{Cai13, Chenu17}.

\subsection{Stochastic Lindblad master equation}

Consider the stochastic Hamiltonian~\cite{Chenu17}
\begin{equation}\label{eq:stoH}
    \hat{H}(t) = \hat{H}_0(t) + \sum_j \lambda_j x_j(t)  \hat{L}_j,
\end{equation}
where $\hat{H}_0(t)$ is the non-stochastic part and
\begin{equation}
    \langle x_j(t) \rangle = 0, \quad \langle x_j(t) x_k(t') \rangle = \delta_{jk} \delta(t - t'). \label{eq:whitenoise}
\end{equation}
Here, $\lambda_j x_j(t)$ represents independent white noise in the strength or phase of the field, where $x_j(t)$ is normalized as in Eq.~\eqref{eq:whitenoise}. For convenience, we consider real-valued fields, i.e., $\lambda_j$ and $x_j(t)$ are real numbers. Then, the density matrix $\rho(t)$, averaged over all realizations of white noise $x_j(t)$, follows the stochastic Lindblad master equation
\begin{equation}
    \frac{d \rho(t)}{dt} = -i [\hat{H}_0, \rho(t)] + \sum_j \lambda^2_j \Big( \hat{L}_j \rho(t) \hat{L}_j^\dagger - \frac{1}{2} \{ \hat{L}_j^\dagger \hat{L}_j, \: \rho(t)\}  + \hat{L}_j^\dagger \rho(t) \hat{L}_j - \frac{1}{2} \{ \hat{L}_j \hat{L}_j^\dagger, \: \rho(t) \} \Big). \label{eq:stochmastereqn}
\end{equation}
where $\hat{L}_j$ and $\hat{L}_j^\dagger$ turn into Lindblad operators. The proof can be found in Ref.~\cite{Chenu17}. Note that the strengths assigned to Lindblad operators $\hat{L}_j$ and $\hat{L}_j^\dagger$ are the same. 

\subsection{Thermal-state preparation using randomized SDK operations} \label{sec:supp_randphase}

First, we simulate the infinite-temperature heating, described by a pair of Lindblad operators $\hat{L}_1 = \sqrt{\Gamma'}\hat{b}$ and $\hat{L}_2 = \sqrt{\Gamma'}\hat{b}^\dagger$, by randomizing the motion phase $\phi_m$ of the SDK operations. The thermal state can be prepared by applying this simulated heating on the motional ground state. The Hamiltonian of the resonant ($\delta_m=0$) SDK operation with sideband Rabi frequency $\tilde{\Omega}$ and stochastically varying $\phi_m(t)$ is given by
\begin{equation} \label{eq:SDKrandphi}
    \hat{H}(t) = \frac{\tilde{\Omega}}{2}(\hat{b}e^{-i \phi_m(t)} + \hat{b}^\dagger e^{i \phi_m(t)}) = \frac{\tilde{\Omega}}{2}\left( \cos\phi_m(t) (\hat{b} + \hat{b}^\dagger) + \sin \phi_m(t) (-i)(\hat{b}-\hat{b}^\dagger)\right), 
\end{equation}
where $\hat{b}$ here is the annihilation operator of the trapped ions' motional mode. The spin part is ignored as the qubit is prepared in the $+1$-eigenstate of the SDK Hamiltonian's spin operator. 

In practice, we approximate white noise in $\phi_m(t)$ by performing $N'$ discrete steps of SDK operations, where $\phi_m$ of each step is a constant value randomly drawn from $[0, 2\pi)$. Thus, the values of $\phi_m(t)$ and $\phi_m(t')$ are equal (completely independent) when times $t$ and $t'$ belong to the same (different) discrete step(s). At the limit $\tau'\rightarrow 0$, where $\tau'$ is the duration of each step,
\begin{equation}
    \langle \cos \phi_m(t) \cos \phi_m(t') \rangle =  \langle \sin \phi_m(t) \sin \phi_m(t') \rangle \rightarrow \frac{\tau'}{4} \delta(t - t').
\end{equation}
As $x_1(t)$ and $x_2(t)$ need to be normalized, we plug in $x_1(t) = (2/\sqrt{\tau'}) \cos\phi_m(t)$, $x_2(t)=(2/\sqrt{\tau'}) \sin\phi_m(t)$, $\lambda_1 = \lambda_2 = \tilde{\Omega} \sqrt{\tau'} /4$, $\hat{L}_1 = \hat{b}+\hat{b}^\dagger$, and $\hat{L}_2 = (-i)(\hat{b}-\hat{b}^\dagger)$ into Eq.~\eqref{eq:stoH} such that, at the limit $\tau'\rightarrow 0$, we obtain the master equation 
\begin{align}
    \frac{d \rho(t)}{dt} = \frac{\tilde{\Omega}^2 \tau'}{8} \Big( 
    &(\hat{b}+\hat{b}^\dagger) \rho(t) (\hat{b}+\hat{b}^\dagger) 
    - \frac{1}{2} \{ (\hat{b}+\hat{b}^\dagger)(\hat{b}+\hat{b}^\dagger), \: \rho(t)\} \nonumber \\ 
    - &(\hat{b}-\hat{b}^\dagger) \rho(t) (\hat{b}-\hat{b}^\dagger)
    + \frac{1}{2} \{ (\hat{b}-\hat{b}^\dagger) (\hat{b}-\hat{b}^\dagger), \: \rho(t) \} 
    \Big) \nonumber\\
    = \frac{\tilde{\Omega}^2 \tau'}{4} \Big(
    & \hat{b}\rho(t)\hat{b}^\dagger - \frac{1}{2} \{ \hat{b}^\dagger\hat{b}, \: \rho(t)\}  + \hat{b}^\dagger \rho(t) \hat{b} - \frac{1}{2} \{ \hat{b}\hat{b}^\dagger, \: \rho(t) \}
    \Big).
\end{align}
This is the Lindblad master equation with Lindblad operators $\hat{L}_1 = \sqrt{\Gamma'}\hat{b}$ and $\hat{L}_2 = \sqrt{\Gamma'}\hat{b}^\dagger$, where
\begin{equation} \label{eq:Gammaprime}
    \Gamma' =  \frac{\tilde{\Omega}^2 \tau'}{4}.
\end{equation}
The heating strength $\Gamma'$ is essentially the rate of increase in phonon number over time. Thus, by applying $N'$ steps of these SDK operations on the motional ground state, we prepare the thermal state with average phonon number $\bar{n} = N'\Gamma'\tau' = N' \tilde{\Omega}^2 \tau'^2/4$ when averaged over many realizations of random phase $\phi_m(t)$. In our experiments we tune $\tau'$ to control $\bar{n}$. 

\subsection{Dephased time evolution using randomized SDK operations} \label{sec:supp_randdetuning}

Next, we simulate the dephasing during time evolution, described by a single Lindblad operator \mbox{$\hat{L}_1 = \sqrt{\Gamma}\hat{b}^\dagger \hat{b}$}, by adding random offsets to the motion detuning $\delta_m$ of the SDK operations. The coherent evolution with respect to $\hat{H}_{SO}$ is simulated using $N$ Trotterization steps, where the SDK operation's duration for each step is $\tau$. Without losing generality, we consider the case where the SDK operations that simulate the coherent evolution use zero motion detuning, such that $\delta_m$ denotes the offset ($\nu \hat{b}^\dagger \hat{b}$ term of $\hat{H}_{SO}$ can be simulated by modulating $\phi_m$ over time, as discussed in the main text). In an appropriate frame, the Hamiltonian of the SDK operation with stochastically varying $\delta_m(t)$ is given by
\begin{equation}
    \hat{H}(t) = \hat{H}_0(t) + \delta_m(t)\hat{b}^\dagger\hat{b},
\end{equation}
where $\hat{H}_0(t)$ is the part that simulates the coherent evolution. 

In our experiments, $\delta_m(t)$ is constant over each Trotterization step, and its value at each step is randomly drawn from a normal distribution $\mathcal{N}(0, \delta_{m,\text{std}})$.  Again, the values of $\delta_m(t)$ and $\delta_m(t')$ are equal (completely independent) when times $t$ and $t'$ belong to the same (different) Trotterization step(s). At the limit $\tau \rightarrow 0$, we obtain
\begin{equation}
    \langle \delta_m(t) \delta_m(t') \rangle \rightarrow \frac{\delta_{m,\text{std}}^2 \tau}{2} \delta(t - t').
\end{equation}
In order to normalize $x_1(t)$, we plug in $x_1(t) = \sqrt{2/\tau}(\delta_m(t)/\delta_{m,\text{std}})$, $\lambda_1 = \sqrt{\tau/2} \delta_{m,\text{std}}$, and $\hat{L}_1 = \hat{b}^\dagger\hat{b}$ into Eq.~\eqref{eq:stoH}, which gives the master equation
\begin{equation}
    \frac{d \rho(t)}{dt} = -i[\hat{H}_0(t), \rho(t)] + \delta_{m,\text{std}}^2 \tau \Big(
    \hat{b}^\dagger\hat{b}\rho(t)\hat{b}^\dagger\hat{b} - \frac{1}{2} \{ \hat{b}^\dagger\hat{b}\hat{b}^\dagger\hat{b}, \: \rho(t)\} 
    \Big),
\end{equation}
which is the Lindblad master equation with \mbox{$\hat{L}_1 = \sqrt{\Gamma}\hat{b}^\dagger \hat{b}$}. We set the standard deviation $\delta_{m,\text{std}}$ such that
\begin{equation}
    \frac{T}{N\tau}\Gamma = \delta_{m,\text{std}}^2 \tau,
\end{equation}
where the left-hand side is rescaled, as we simulate the evolution up to time $T$ using SDK operations of total duration $N\tau$. This completes our method of simulating the time evolution of the dephased spin-oscillator model. 

\subsection{Errors due to finite discretization}\label{finitesize}

As we approximate white noise into discrete steps where randomly drawn parameter is assigned to each step, errors may occur due to finite duration of each step. Here we numerically analyze the errors due to finite discretization. 

\begin{figure*}[ht!]
\centering
\includegraphics[width=\textwidth]{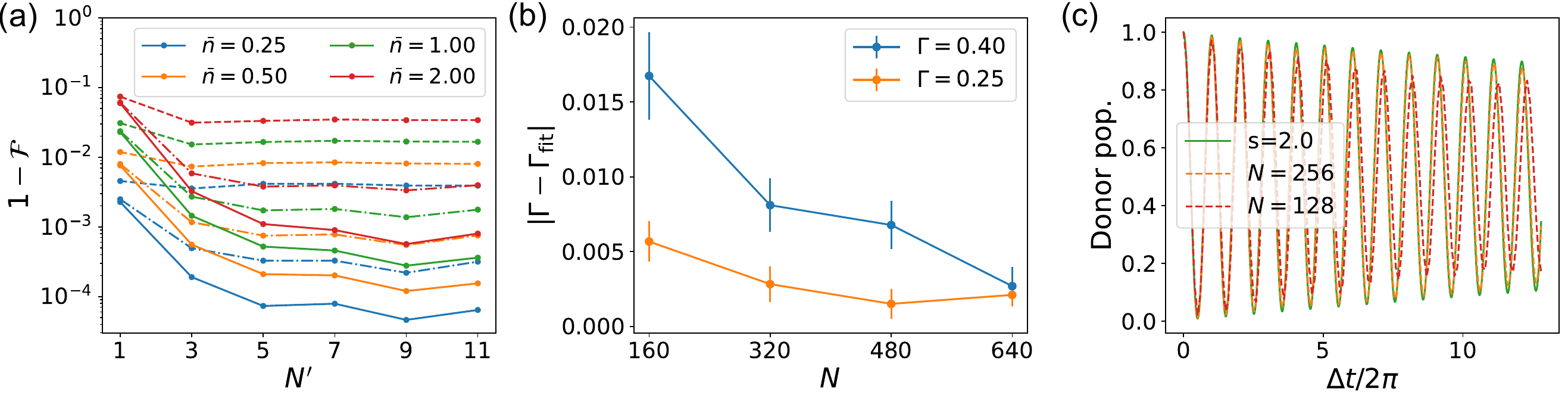}
\caption{\textbf{Errors due to finite discretization.} \textbf{(a)} Errors $1-\mathcal{F}$ in the thermal state prepared using $N'$ steps of SDK operations, where a random phase is assigned to each step. $\mathcal{F}$ is the fidelity between the thermal state of average phonon number $\bar{n}$ and the ensemble of prepared states. The ensemble is obtained by averaging over 20 (dashed), 200 (dot-dashed), and 2000 (solid) trials. For the experimental data in Fig.~2a of main text, we use 20 trials and $N'=5$. \textbf{(b)} Errors in the fitted values of $\Gamma$, where the population curve obtained by SDK operations is fitted to that of the dephased spin-oscillator model as in Fig.~2e of main text. $N$ is the number of Trotterization steps, each assigned a random detuning. Same Hamiltonian parameters are used as in Fig.~2b of main text. The population curve is averaged over 2000 random trials. Error bars represent the uncertainty of the fitted $\Gamma$ due to shot noise, where we assume the qubit population is measured only once for each random trial. For the experimental data in Fig.~2b of main text, we use $N=160$, and repeat each of 20 random trials 100 times. \textbf{(c)} Simulations of the experiment in Fig.~4d of main text. Green solid curve represents the theoretical predictions of the spin-boson model with spectral density given by the green solid curve in Fig. 4a of main text. Orange (red) dashed curve is the expected results derived from simulating $N=256$ (128) Trotterization steps of SDK operations with random detuning. For the experimental data in Fig.~3d, we use $N=128$, which closely match the red dashed curve as expected.}\label{fig_supp_disc}
\end{figure*}

First, we study the error in thermal-state preparation using randomized operations. In Sec.~\ref{sec:supp_randphase}, we show that SDK operations with random phase simulate infinite-temperature heating, and therefore prepare the thermal state (when starting from the ground state), in the limit $\tau' \rightarrow 0$. With a fixed target average phonon number $\bar{n}$ of the thermal state, this limit corresponds to $N' \rightarrow \infty$. 

We numerically simulate the ensemble of states obtained by many trials of SDK operations with random phase, and plot the fidelity $\mathcal{F}$ between the ensemble and the target thermal state. Figure~\ref{fig_supp_disc}a shows that the thermal state can be prepared with reasonably high fidelity using only a few ($N'<10$) steps, for the range of $\bar{n}$ considered in this paper. In particular, using 2000 trials and $N'=9$ discrete steps, the thermal state with $\bar{n}=2$ can be prepared with error $1-\mathcal{F} < 10^{-3}$. 

For all values of $\bar{n}$, as $N'$ increases from 1, the error decreases at first but then stops decreasing. This shows that in practice, setting $N'$ to a large number may not be effective in preparing a high-fidelity thermal state. Instead, the error floor can be lowered by increasing the number of trials, as shown in Fig.~\ref{fig_supp_disc}a. Also note that for a fixed target fidelity, preparing the thermal state with larger $\bar{n}$ requires more resources (larger $N'$ and/or more trials). 

Next, we analyze the error in the rate of dephasing simulated using randomized operations. In Sec.~\ref{sec:supp_randdetuning}, we show that adding random detunings to the SDK operations is equivalent to dephasing during the time evolution, in the limit $\tau \rightarrow 0$, or equivalently, $N \rightarrow \infty$. 

As in Fig.~2b and e of main text, we numerically simulate the SDK operations with random detuning and fit the population curve to that of the dephased spin-oscillator model, where the dephasing rate $\Gamma$ is used as the fitting parameter. We choose to use the underdamped cases ($\Gamma=0.25$ and $0.4$), as coherent oscillations give more reliable fitting results. To evaluate the effects of finite $N$, we plot the errors in the fitted values of $\Gamma$ for various values of $N$ in Fig.~\ref{fig_supp_disc}b. As expected, the error decreases as $N$ increases. 

Note that the error due to finite $N$ discussed here is different from the uncertainty plotted in Fig.~2e of main text, which is due to finite number of random trials. For a larger number (e.g. 2000) of trials, the uncertainty is significantly smaller than the error due to finite $N$. However, this uncertainty does not take into account shot noise, i.e., uncertainty in qubit population due to finite number of repetitions of binary measurements. For performing a large number of trials, repeating each trial many times to reduce shot noise may require prohibitively long experiment time. 

To evaluate the effects of shot noise, we simulate performing 2000 random trials where the qubit state is measured only once for each trial and time step. Specifically, for each random trial, we add a number (0 or 1) randomly drawn from the binomial distribution of success probability $p$, where $p$ is the simulated qubit population for the corresponding trial and time step. The sum of these numbers over all trials, divided by the number of trials, is the recorded population for each time step. The population curve obtained in this way is fitted as described above. We repeat this entire procedure 100 times and obtain the standard deviation of the fitted $\Gamma$ values, reported as error bars in Fig.~\ref{fig_supp_disc}b. For $N \leq 480$, the magnitude of this uncertainty due to shot noise is significantly smaller than that of the error in fitted $\Gamma$ due to finite $N$. Therefore, higher accuracy in $\Gamma$ can be achieved with a fixed total number of measurements by performing a larger number of random trials with fewer repetitions. This requires a faster control software that compiles a given set of control parameters~\cite{Dalvi23}.

Finally, we discuss the errors due to finite discretization in Fig.~4d of main text. The experimental data obtained using $N=128$ Trotterization steps deviate non-negligibly from the theoretical predictions of the simulated spin-boson model. To explain this mismatch, we numerically simulate performing $N$ SDK operations with random detuning, for $N=128$ and 256. Figure~\ref{fig_supp_disc}c shows that the results for $N=256$ match with the theoretical predictions significantly better than the results for $N=128$. Thus, we expect that the mismatch of our quantum simulation results in Fig.~4d of main text can be removed by using a larger number of Trotterization steps $N$. 

In the experiment, we choose $N=128$ steps owing to the memory constraints of the controlling software. The RF signals utilized for modulating the laser pulses are compiled within a local CPU and stored in the block RAM (BRAM) of a field-programmable gate array (FPGA) board, which has restricted storage capacity for a lengthy instruction of operations. Upgrading both the controlling software and hardware could enable us to perform quantum simulations with a larger number of Trotterization steps.

\section{Methods of fitting spectral densities}

Accurately modeling the dynamics of the spin-boson model requires a reliable representation of the spectral density function $J(\omega)$, which encapsulates the frequency-dependent coupling between the spin and the bosonic environment. In this work, we employ a fitting strategy that approximates the spectral density using a linear combination of basis functions (with some linear constraints, see below), enabling trapped-ion simulations that use finite number of oscillators to simulate a continuum of environment modes.
The spectral density to be approximated is from Leggett et al.~\cite{Leggett87}, given by
\begin{align}
    J_{\rm Legg}(\omega) = A \omega^s \omega_c^{1-s} e^{-\omega / \omega_c}.
\end{align}

There are two approaches to fit a spectral density, the spectral density based method and the correlation function based method. The spectral density based method is used to obtain the fitting in the main text. The correlation function based method may serve as an alternative when more flexibility in the fitting parameters is allowed, and we include it here for the readers to better understand the linear constraints used in the spectral density based method.

\subsection{Spectral density based method}
The first step in our approach is the discretization of the spectral density function. We utilize the Legendre discretization method \cite{de2015discretize} to obtain a set of discretized frequencies $\omega_n$ and the corresponding weights $w_n$ based on the roots and weights of the Legendre polynomials. The number of grid points $N$ and the frequency domain $[\omega_l, \omega_r]$ ($\omega_l \geqslant 0$ and $\omega_r < \infty$) are chosen to ensure sufficient resolution and coverage of the relevant frequency range. As discussed in the main text, $\omega_l$ and $\omega_r$ can be chosen to be values such that $[\omega_l, \omega_r]$ cover the range of on-resonance frequencies without detrimenting the correctness of the spin dynamics.

Next, we evaluate the spectral density function $J(\omega)$ at the discretized frequencies $\omega_n$ and calculate the coupling strengths $V_n^2$ as the product of $J(\omega_n)$ and the corresponding weights $w_n$:
\begin{align}
    V_n^2 = \frac{J_{\rm Legg}(\omega_n)}{\pi} w_n.
    \label{supp:eq:ohmic}
\end{align}
These coupling strengths represent the discretized version of the spectral density and serve as the target values for the fitting process.

To approximate the spectral density, we employ a sum of Lorentzian basis functions, which provide a flexible and physically motivated representation~\cite{tannor1999}. The approximated spectral density function $J_{\text{approx}}(\omega)$ is given by
\begin{gather}
    J_{\mathrm{approx}}(\omega) = \sum_{m=1}^{M} J_{\mathrm{Lo},m}(\omega), \\
    J_{\mathrm{Lo},m}(\omega) = \kappa_m^2 \left(\frac{\Gamma_m/2}{(\Gamma_m/2)^2 + (\omega - \nu_{m})^2} - \frac{\Gamma_m/2}{(\Gamma_m/2)^2 + (\omega + \nu_{m})^2}\right),
    \label{supp:eq:lorentzian}
\end{gather}
where $\kappa_m$, $\gamma_m$, and $\nu_{m}$ are the parameters of the $m$-th Lorentzian basis function, representing the coupling strength, damping rate, and peak frequency, respectively. These parameters are optimized during the fitting process to minimize the difference between the target spectral density and the approximating function. As what is done for Leggett's spectral density in Eq.~\eqref{supp:eq:ohmic}, the approximating function $J_\text{approx}(\omega)$ is also discretized using the Legendre discretization method, resulting in the same set of frequencies $\{\omega_n\}$ and weights $\{w_n\}$ but different coupling strengths $\tilde{V}_n^2$:
\begin{align}
    \tilde{V}_n^2 = \frac{J_\text{approx}(\omega)}{\pi} w_n
\end{align}

The optimization is performed by minimizing an objective function defined as the sum of the relative differences between the target and approximated spectral densities at the discretized frequencies:
\begin{align}
    \text{objective} = \sum_{n=1}^{N} \frac{\left|V^2_n - \tilde{V}^2_n \right|}{|V^2_n|}
\end{align}
We employ Bitmask Evolution Optimization (BITEOPT), a powerful black-box optimization algorithm, to find the optimal values of the Lorentzian parameters. The specific implementation of the algorithm used can be found in Ref. \citenum{fcmaes2022}.

To ensure physically reasonable values and avoid the deviation of the behaviour of the Lindblad oscillators from the Lorentzian spectral density in Eq.~\eqref{supp:eq:lorentzian}, we impose constraints on the Lorentzian parameters during the optimization process. Specifically, we introduce a linear constraint that relates the damping rate $\gamma_m$ and the peak frequency $\nu_{m}$ of each Lorentzian basis function:
\begin{align}
    \gamma_m \leq \frac{\nu_{m}}{2}.
\end{align}
This constraint is motivated by the condition under which a Lindblad oscillator can be approximated by a Lorentzian spectral density [see the discussion below Eq.~\eqref{eq:CtJomega}]. Additionally, bounds are placed on the parameters to restrict them to physically meaningful ranges, which helps reduce the size of the optimization space. For example, the range of on-resonance frequencies $[\omega_l,\omega_r]$ is set based on the coupling of the spin system. Additionally, the Lorentzian coupling strengths $\kappa_m$ for each component are restricted to values smaller than the overall system-bath coupling strength.

\subsection{Correlation function based method}
Instead of minimizing the discrepancy between spectral densities, 
we can compare the bath correlation functions (BCF) obtained from the target spectral density (SD-BCF) and the BCF of a Lindblad oscillator (LD-BCF), since the dynamics of the spin in the spin-boson model is solely determined by the BCF~\cite{Mascherpa20,somoza2019,Tamascelli18,Tamascelli19,Lemmer18}. The BCFs are calculated using the following expressions~\cite{Lemmer18}
%[the BCF of a Lorentzian spectral density (SDL-BCF) that has the same coupling strength, damping rate, and peak frequency as the Lindblad oscillator is also given for comparison]:
\begin{align}
    C_{\text{SD-BCF}}(t) & = \int_0^\infty d\omega \frac{J_{\text{Legg}}(\omega)}{\pi} \left(\frac{\cos \omega t }{\tanh({\beta \omega}/{2})} - i \sin \omega t \right) \\ %\approx \sum_{n=1}^{N} V_n^2 \left(\frac{\cos(\omega_n t)}{\tanh(\beta\omega_n/2)} - i \sin(\omega_n t)\right) \\
    C_{\text{LD-BCF}}(t) & = \kappa^2 e^{-\Gamma t/2} \left(\frac{\cos \nu t}{\tanh({\beta\nu}/{2})} - i \sin \nu t \right) =  \int_0^\infty \frac{\mathrm{d}\omega }{\pi} \left(\frac{\tilde{J}_{\mathrm{Lo}}(\beta,\omega)\cos \omega t }{\tanh({\beta \omega}/{2})} - i 
    J_{\mathrm{Lo}}(\omega)\sin \omega t \right)\\
    J_{\mathrm{Lo}}(\omega) & = \kappa^2 \left(\frac{\Gamma/2}{(\Gamma/2)^2 + (\omega - \nu)^2} - \frac{\Gamma/2}{(\Gamma/2)^2 + (\omega + \nu)^2}\right) \\
    \tilde{J}_{\mathrm{Lo}}(\beta,\omega) & = \kappa^2 \frac{\tanh(\beta\omega/2)}{\tanh(\beta\nu/2)}\left(\frac{\Gamma/2}{(\Gamma/2)^2 + (\omega - \nu)^2} - \frac{\Gamma/2}{(\Gamma/2)^2 + (\omega + \nu)^2}\right) 
    % \\ 
    % C_{\text{SDL-BCF}} & = \int_0^\infty d\omega \frac{J(\omega)}{\pi} \left(\frac{\cos \omega t }{\tanh({\beta \omega}/{2})} - i \sin \omega t \right) 
\end{align}
where $J_{\mathrm{Lo}}$ is the Lorentzian spectral density and $\tilde{J}_{\mathrm{Lo}}$ is the Lorentzian spectral density modified by a factor $\frac{\tanh(\beta\omega/2)}{\tanh(\beta\nu/2)}$.

Fitting the target correlation function $C_{\text{SD-BCF}}(t)$ at a target temperature $\beta$ by a linear combination of Lindblad oscillators at temperature ${\beta_m}$ amounts to matching the real and imaginary parts inside the integral over $\omega$, simultaneously. By allowing the variation of the temperatures of the Lindblad oscillators, we obtain the following two approximation formats, corresponding to the real and imaginary parts of the correlations functions, respectively:
\begin{equation}
\begin{aligned}
    \text{Real part:} & \quad \frac{J_{\text{Legg}}(\omega)}{\tanh(\beta\omega/2)}  \approx \sum_m^M \frac{\tilde{J}_{\mathrm{Lo},m}(\beta_m,\omega)}{\tanh(\beta_m\omega/2)}  \Longrightarrow {J_{\text{Legg}}(\omega)} \approx \sum_m^M \frac{\tanh(\beta\omega/2)}{\tanh(\beta_m\nu/2)} {J}_{\mathrm{Lo},m}(\omega) \\
    \text{Imaginary part:} & \quad J_{\text{Legg}}(\omega)  \approx \sum_m^M {J}_{\mathrm{Lo},m}(\omega). 
    \label{supp:eq:approximants}
\end{aligned}
\end{equation}

The two conditions can be fulfilled simultaneously in principle, provided enough flexibility of the parameters $\beta_m$, $\kappa_m$, $\Gamma_m$, $\nu_m$ and the number of Lindblad oscillators $M$. By using the approximations in Eq.~\eqref{supp:eq:approximants}, we convert the task of fitting the correlation function into a task of fitting the spectral density with two approximations simultaneously. The advantage of the conversion is that, once again, we can specify an on-resonance vibrational frequency range $[\omega_l, \omega_r]$, instead of working on the full frequency range $[\omega_l=0,\omega_r=\infty]$.

In summary, the proposed two fitting strategies provide a systematic and flexible approach to approximate complex spectral density functions using a linear combination of Lorentzian/Lindblad oscillators. The discretization of the spectral density ensures a tractable computational framework, while the optimization procedure minimizes the discrepancy between the target and approximated functions. The incorporation of physically motivated constraints on the Lorentzian parameters helps to maintain the physical relevance of the fitted spectral density.

\section{Leggett's spectral densities with fixed values at resonance}

In Fig. 4 of main text, we simulate Leggett's spin-boson models $J_{\rm Legg}(\omega) = A \omega^s \omega_c^{1-s} e^{-\omega / \omega_c}$ in the weak spin-bath coupling regime using spectral densities composed of two Lorentzian peaks. As $s$ is increased, the spectral density value at resonance  $J_{\rm Legg}(\omega=\Delta)$ decreases, and thus the spin-state population's oscillations are damped more weakly. This qualitatively agrees with the analysis of Leggett et al.~\cite{Leggett87} that the spin dynamics exhibit a transition from localization (extremely strong damping) or incoherent relaxation to coherent oscillations (weak damping) as the exponent $s$ is increased from sub-Ohmic ($s<1$) to Ohmic ($s=1$) to super-Ohmic ($s>1$) values in the strong spin-bath coupling regime.

Meanwhile, another interesting question is how the spin dynamics behavior of Leggett's model changes when the exponent $s$ is varied under \textit{fixed} spectral density value at resonance $J_{\rm Legg}(\omega=\Delta)$. This enables isolating the effects of the spectral density's slope (to first order) while fixing the zero-th order contribution from the resonance value. 

\begin{figure*}[ht!]
\centering
\includegraphics[width=\textwidth]{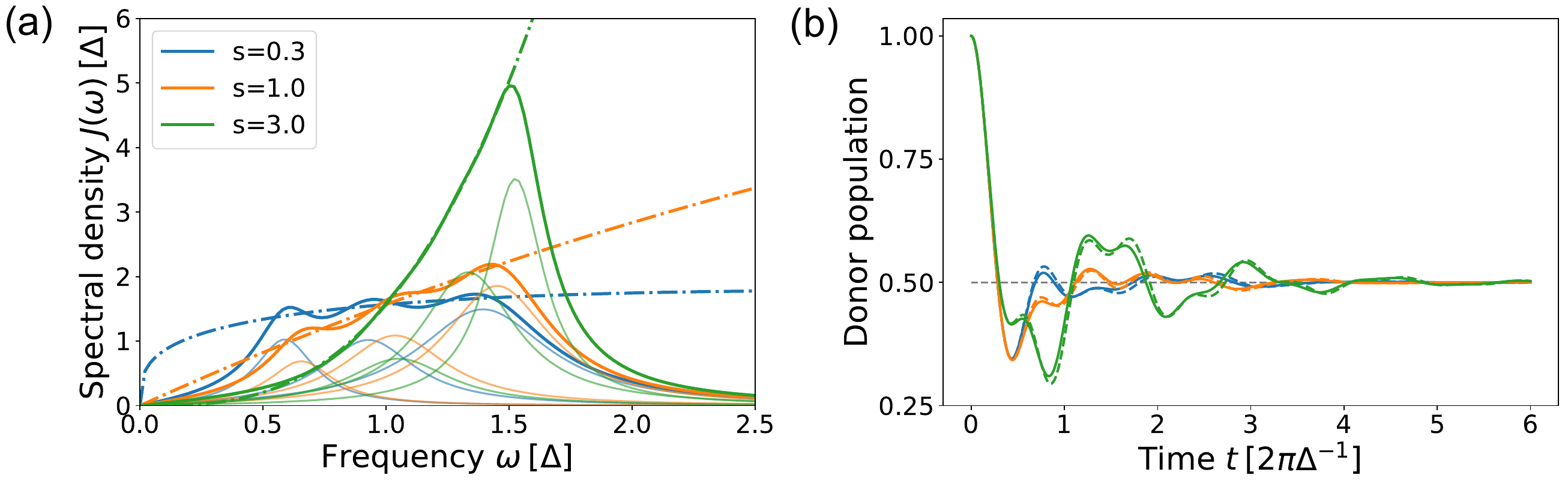}
\caption{\textbf{Numerical simulations based on Leggett's spectral densities with fixed values at resonance. (a)} $J_{\rm Legg}(\omega)$ (dash-dotted) and the fitted spectral density consisting of three Lorentzian peaks (solid) for $s=0.3$, 1, and 3. The constituent Lorentzian peaks are shown in faint solid curves. \textbf{(b)} Dynamics of the dephased spin-oscillator models (solid) and the spin-boson models (dashed) with spectral densities given by the solid curves in (a).  }\label{fig_supp_Leggett}
\end{figure*}

Indeed, our method is capable of capturing the difference in the spin dynamics of sub-Ohmic, Ohmic, and super-Ohmic baths even when the resonance value is fixed. Instead of fixing the value of $A$ as in the main text, we consider fixing $J_{\rm Legg}(\omega=\Delta)$ as $s$ is varied by setting the value of $A$ accordingly ($\omega_c=10\Delta$ is fixed), and fit the spectral density composed of 3 Lorentzian peaks to $J_{\rm Legg}(\omega)$ for each $s$. We find reasonably good fitting results by setting the target bandwidth as $\omega \in [0.5\Delta, 1.5\Delta]$ for the cost function of fitting, which is significantly broader than $[0.9\Delta, 1.1\Delta]$ used in the main text. The target and fitted spectral densities are shown in Fig.~\ref{fig_supp_Leggett}(a). Note that in this example, the resonant value $J_{\rm Legg}(\omega=\Delta)$ of the target spectral densities is fixed to $1.57\Delta$, which is in the relatively strong spin-bath coupling regime. 

Figure~\ref{fig_supp_Leggett}(b) shows the spin dynamics of the dephased spin-oscillator models and the spin-boson models with the 3-peak spectral densities given in Fig.~\ref{fig_supp_Leggett}(a). In the super-Ohmic case ($s=3$), the donor-state population undergoes at least two full oscillations before the population decays to 0.5 (with negligible oscillation amplitude), whereas in the Ohmic ($s=1$) and sub-Ohmic ($s=0.3$) cases the population barely undergoes a single oscillation before decaying to near 0.5. This shows that strong coupling to low-frequency bath oscillators induces more rapid population decay than strong coupling to high-frequency bath oscillators, as lower-frequency oscillators are excited to higher phonon-number states and therefore are more strongly subject to dephasing ($\because \hat{a}^\dagger \hat{a} \ket{n} = n \ket{n}$) or damping ($\because \hat{a} \ket{n} = \sqrt{n} \ket{n-1}$). Also, note that the spin-bath coupling is sufficiently strong for the spin-boson models with different values of $s$ to exhibit different dynamics behavior, yet not too strong such that population dynamics of the spin-boson models do not deviate significantly from those of the dephased spin-oscillator models. Thus, our method of quantum simulations is capable of distinguishing the effects of the exponent $s$ of Leggett's spectral density even when the resonance value is fixed. 

Comparing the Ohmic ($s=1$) and sub-Ohmic ($s=0.3$) cases, the sub-Ohmic spin-boson model exhibits a faster population decay to near 0.5. This is inconsistent with the simulation results in Fig.~3 of Ref.~\cite{zhang2021efficient}, where the donor-state population decay is significantly faster in the Ohmic case. The analysis of Ref.~\cite{zhang2021efficient} is different from ours in several ways; (i) the reorganization energy $\int d\omega J(\omega)/\omega$ is fixed among different $s$ values, rather than the resonant $J(\omega=\Delta)$ value, (ii) the spin Hamiltonian consists of four states, and (iii) a stronger spin-bath coupling is considered. Future works may identify in which conditions the population dynamics exhibit ``critical damping'' (fastest population decay) with the Ohmic spectral density. Quantum simulations using a larger number of oscillators would be able to cover a wider bandwidth and capture the effects of strong coupling to a broad band of high-frequency oscillators.

\bibliography{supplementary}% common bib file